\documentclass[aps,prd,preprint,preprintnumbers,showpacs,nofootinbib]{revtex4-1}
\usepackage{amsmath,amsfonts,amssymb}
\usepackage{graphicx}
\usepackage{multirow}
\usepackage{bigstrut}
\usepackage{makecell}
\usepackage{xcolor}
\usepackage[left=2.5cm,right=2.5cm]{geometry}

\begin{document}
\title{Hyperon resonances coupled to pseudoscalar- and vector-baryon channels}
\author{K.~P.~Khemchandani$^{1}$\footnote{kanchan.khemchandani@unifesp.br}}
\author{A.~Mart\'inez~Torres$^{2}$\footnote{amartine@if.usp.br}}
\author{ J.~A.~Oller$^3$\footnote{oller@um.es}}

 \affiliation{
$^1$ Departamento de Ci\^encias Exatas e da Terra, Universidade Federal de S\~ao Paulo, Campus Diadema, Rua Prof. Artur Riedel, 275, Jd. Eldorado, 09972-270, Diadema, SP, Brazil. \\
$^2$ Instituto de F\'isica, Universidade de S\~ao Paulo, C.P 66318, 05314-970 S\~ao Paulo, SP, Brazil.\\
$^3$ Departamento de F\`isica, Universidad de Murcia, E-30071 Murcia, Spain.
}
\begin{abstract}
We study hyperon resonances by solving coupled channel scattering equations. The coupled systems include pseudoscalar- and vector-baryon channels. The parameters of the model are restricted by making a $\chi^2$-fit to the cross section data on processes: $K^- p \to K^- p$, $K^- p \to \bar K^0 n$, $K^- p \to \eta \Lambda$, $K^- p \to \pi^0 \Lambda$, $K^- p \to \pi^0 \Sigma^0$, $K^- p \to \pi^\pm \Sigma^\mp$. Data on the energy level shift and width of the $1s$ state of the kaonic hydrogen, as well as some cross-section ratios near the threshold are also considered in the fit. Two types of fits are found as a result. In both cases, the properties of $\Lambda(1405)$ are well reproduced. In addition to this, a $\Sigma$ state is also found with mass around 1400 MeV. Cross sections, obtained with one of the two fits, are found to stay close to the data at energies away from the thresholds too, and as a result resonances with higher masses have also been studied.
\end{abstract}

\maketitle
\section{Introduction}
Investigating low-energy meson-baryon interaction, with nonzero strange quantum number, is of great importance to several interrelated topics in nuclear and hadron physics, such as the determination of the nature of the low-lying hyperons \cite{Kaiser:1995eg, osetramos, Oller:2000fj, Jido:2003cb, Hyodo:2011ur, Lu:2013nza, Kamano:2014zba, Garcia-Recio:2015jsa, Kamano:2016djv, Ramos:2016odk}, the existence of kaonic-nuclear bound states, which has motivated several experiments \cite{Skurzok:2018bur,Dote:2017lyy,Sakuma:2017muh}, studies of kaon producing reactions which are, in turn, useful to understand the interactions of kaons in a dense medium \cite{Cabrera:2015bga}, etc. The key motivational idea behind several related works is that the strangeness $-1$ meson-baryon interaction is attractive in nature, and it is especially interesting in the $s$-wave since, as now widely accepted, it generates the isoscalar resonance $\Lambda(1405)$. The list of references on this topic is extensive, but for some of the recent works we refer the reader to Refs.~\cite{Mai:2018rjx,Miyahara:2018onh,Roca:2017wfo,Wang:2017rjs,Liu:2016wxq,Kamiya:2016jqc,Molina:2015uqp,Oset:2015ksa,Torres:2013boa}. A lot of effort is being put in by the lattice community too, bringing valuable information on the topic \cite{Hall:2014uca,Menadue:2011pd,Ishii:2007ym,Takahashi:2010nj,Hall:2016kou,Gubler:2016viv,Briceno:2017max}.

There exist evidences for the presence of an isovector  resonance too in nature, with its origin lying in the meson-baryon dynamics, with a mass similar  to $\Lambda(1405)$~\cite{Oller:2000fj,Guo,Wu:2009tu,Wu:2009nw,Gao:2010hy,Xie:2014zga,Xie:2017xwx,Khemchandani:2012ur}. However, the case is less studied, as compared to $\Lambda(1405)$, and the properties of the low-lying $1/2^-$ $\Sigma$(s) obtained from different works are different.  In Ref.~\cite{Oller:2000fj}, a coupled channel study of pseudoscalar-baryon systems was made using a kernel arising from $s$-, $u$-channel exchange of the lightest octet baryon and a contact interaction obtained from the lowest order chiral Lagrangian. The subtraction constant required to calculate the loop function were constrained by fitting relevant data available, namely, the $K^- p \to \bar KN, \pi \Sigma, \pi \Lambda$ cross sections and different cross-section ratios among these processes at the $K^-p$ threshold, as well as the $\pi^+ \Sigma^-$ mass distribution.  As a result, in the case of isospin 1, two $\Sigma$ states were found near the $\bar K N$ threshold: $1440 - i 70$ MeV and $1420 - i 42$ MeV.  The work was further extended by considering next-to-leading-order contributions from the chiral Lagrangian~\cite{Guo} and including data on the energy shift and width of the 1$s$-state in kaonic hydrogen, cross sections on $K^- p \to \eta \Lambda, \pi^0 \pi^0 \Sigma$, etc. In this latter work, the preferred Fit II gives rise to two poles with isospin 1 around the $\bar KN$ threshold with pole positions: $1376 - i 33$  MeV and $1414 - i 12$ MeV. There is another fit to data in Ref.~\cite{Guo}, called Fit~I, with no isospin 1 poles but it is disfavored by the photoproduction data of CLAS~\cite{Lu:2013nza}, because the two poles associated with $\Lambda(1405)$ are both clearly above 1.4~GeV. Independent studies of Refs.~\cite{Wu:2009tu,Wu:2009nw,Gao:2010hy,Xie:2014zga,Xie:2017xwx}  seem to accumulate evidences for a $J^\pi = 1/2^-$  $\Sigma$ with a pentaquark nature, with mass and width 1380 and 60 MeV, respectively, by studying processes different to those considered in Refs.~\cite{Oller:2000fj,Guo}, like: $K^- p \to \Lambda \pi^+ \pi^-$, $\gamma N \to  K^+ \pi \Lambda$, $\Lambda p \to \Lambda p \pi^0$, $\Lambda^+_c \to \eta \pi^+ \Lambda$. In addition to these works, the best fit to the data on $\gamma + p \to K^+ + \Sigma^{\pm,0} + \pi^{\mp,0}$~\cite{Moriya:2013eb} required inclusion of two $1/2^-$ states in the isospin one: $(1413\pm10) - i (26 \pm 5)$ MeV and  $(1394\pm20) - i (75 \pm 20)$ MeV. However, a recent partial-wave analysis ($s$- and $p$-wave) of $S=-1$ low-energy data, including differential cross sections (although it only considers pseudoscalar-baryon contact interactions), does not report finding of any $1/2^-$ $\Sigma$ around 1400 MeV~\cite{Sadasivan:2018jig}. In Ref.~\cite{Mai:2014xna} too, a study of strangeness $-1$ coupled systems has been made including constraints from the CLAS photoproduction data~\cite{Moriya:2013eb} but the discussions made are focused on isospin zero states.
A different analysis of the photoproduction data, consistent with chiral dynamics and unitarity in coupled channels, is conducted in Ref.~\cite{Roca:2013cca} and a $\Sigma^*$ state appears as a strong cusp around the $\bar K N$ threshold, very similar to the $a_0(980)$ shape around the $K\bar K$ threshold. In the present scenario, it is not clear if an isospin one partner of $\Lambda(1405)$ exists, and if it does, it is not clear if  it corresponds to one or two close lying poles in the complex plane. 

Interestingly, in the previous study of $S=-1$ systems \cite{Khemchandani:2012ur}, two  isospin 1 poles were found, though they lied deep in the complex plane, arising from coupled channel meson baryon dynamics (at $1427 - i145$ MeV, $1438 - i198$ MeV). However, the motivation of the work~\cite{Khemchandani:2012ur},  done by two of the present authors, was to build the formalism to couple pseudoscalar- and vector-baryon systems, and it was beyond the scope of Ref.~\cite{Khemchandani:2012ur} to test if the resulting amplitudes reproduced different relevant data.  Nonetheless, the poles of the well studied $\Lambda(1405)$ were reproduced in agreement with  other works. Besides,  the kernels for the pseudoscalar-baryon (PB) systems in Ref.~\cite{Khemchandani:2012ur} were obtained from the contact interaction (the Weinberg-Tomozawa term) coming from the lowest order chiral Lagrangian and the vector-baryon (VB) interactions were calculated by evaluating $s$-, $t$- and $u$-channel diagrams and a contact interaction.  The purpose of our present work is to improve the model used in Ref.~\cite{Khemchandani:2012ur} by including the $s$-, and $u$-channel $1/2^+$ octet baryon-exchange diagrams to the kernels of the pseudoscalar-baryon, which have been found to play an important role  in the generation of $\Sigma$ poles around 1400 MeV in Ref.~\cite{Oller:2000fj}. The importance of these diagrams has been pointed out in other works too, like in Ref.~\cite{Ramos:2016odk}, near the $K\Xi$ threshold. The main motivation of our work is, thus, improving the model of  Ref.~\cite{Khemchandani:2012ur} and to study the existence of light isospin one resonances, those in agreement with the ones predicted in Refs.~\cite{Moriya:2013eb,Oller:2000fj,Guo,Wu:2009tu,Wu:2009nw,Gao:2010hy,Xie:2014zga,Xie:2017xwx}. However, the explicit treatment within coupled channels of the vector-baryon interactions gives rise to higher-order contributions beyond next-to-leading (NLO) chiral perturbation theory ($\chi$PT). From this point of view, our study can also be seen as a partial check of the stability of the unitarized NLO $\chi$PT results \cite{Guo,Mai:2014xna}. 

With the improved PB kernels, we constrain the parameters of the formalism (mainly the subtraction constants required to calculate the loop functions),  to reproduce different available experimental data and test if the low lying $\Sigma$s found in Ref.~\cite{Khemchandani:2012ur} move closer to the real axis, and could correspond to the $\Sigma$s found in Refs.~\cite{Oller:2000fj,Guo,Wu:2009tu,Wu:2009nw,Gao:2010hy,Xie:2014zga,Xie:2017xwx}. The generation of the states like $\Lambda(1405)$ or $\Sigma$ with a similar mass is not expected to get important contributions from VB dynamics, but the inclusion of VB dynamics in the model can be very relevant in determining useful informations. For example, with our model we can obtain the R-VB couplings (where R is a resonance, like $\Lambda(1405)$, $\Lambda(1670)$, etc.), which  are required in the calculations of $t$-channel diagrams, with a vector exchange, for processes like the photoproduction/electroproduction of $\Lambda(1405)$.  Additionally, with the improved PB kernels and constrained PB amplitudes, we can obtain more reliable information on the properties of the hyperon resonances arising from the vector-baryon dynamics as well.  

The manuscript is organized as follows. In Sec.~\ref{sec:formalism} we discuss the  Lagrangians from which the meson-baryon interactions are obtained and used as kernels to study nonperturbative scattering in the systems. Toward the end of the same section, we discuss  the idea of carrying out a $\chi^2$-fit, the parameters of the fit, and the data to be considered in the fit. In Sec.~\ref{results} we discuss the details on the results of the fits obtained.  The properties of the resonances found in our study are also given in Sec.~\ref{results}, by categorizing them in different subsections on the basis of their spins and isospins. Finally, we present a summary of the work.

\section{Formalism}\label{sec:formalism}
The problem of hadron scattering gets typically more and more complex as the energy region to be scanned involves opening of more and more thresholds to possible coupled channels. To study hyperon resonances arising from hadron dynamics, with mass up to about  2 GeV, we implement a nonperturbative unitarization method by treating crossed-channel dynamics perturbatively as developed in Refs.~\cite{Oller:1998zr,Oller:1999me,Oller:2000fj}. There is a connection with this method and solving the Bethe-Salpeter equation for contact interactions~\cite{Oller:1997ti,osetramos}. We take into account pseudoscalar- and vector-baryon channels, motivated by the fact that the thresholds of these channels are spread over the energy ranging from 1.25-2.2 GeV, and some of them lie close enough to couple to each other, for example $K \Xi$, $\bar K^* N$. The pseudoscalar meson-baryon interaction diagrams are deduced from the lowest order, $\mathcal O(p)$, Lagrangian~\cite{Meissner:1993ah,ecker,pich,Kaiser:1995eg,Oller:2000fj,osetramos,Oller:2006yh},
\begin{equation}
\mathcal{L}_{PB} = \langle \bar B i \gamma^\mu \partial_\mu B  + \bar B i \gamma^\mu[ \Gamma_\mu, B] \rangle - M_{B} \langle \bar B B \rangle
+  \frac{1}{2} D^\prime \langle \bar B \gamma^\mu \gamma_5 \{ u_\mu, B \} \rangle + \frac{1}{2} F^\prime \langle \bar B \gamma^\mu \gamma_5 [ u_\mu, B ] \rangle,\label{LPB}
\end{equation}
where  $u_\mu = i u^\dagger \partial_\mu U u^\dagger$, and
\begin{eqnarray}
\Gamma_\mu &=& \frac{1}{2} \left( u^\dagger \partial_\mu u + u \partial_\mu u^\dagger  \right), 
\, U=u^2 = {\textrm exp} \left(i \frac{P}{f_P}\right),\label{gammau}
\end{eqnarray}
 with $f_P$ representing the pseudoscalar decay constant, and $P$ ($B$) denoting the matrices of the octet meson (baryon) fields:
\begin{eqnarray} \nonumber
&&P =
\left( \begin{array}{ccc}
\pi^0 + \frac{1}{\sqrt{3}}\eta & \sqrt{2}\pi^+ & \sqrt{2}K^{+}\\
\sqrt{2}\pi^-& -\pi^0 + \frac{1}{\sqrt{3}}\eta & \sqrt{2}K^{0}\\
\sqrt{2}K^{-} &\sqrt{2}\bar{K}^{0} & \frac{-2 }{\sqrt{3}} \eta
\end{array}\right),~~~~
B =
\left( \begin{array}{ccc}
 \frac{1}{\sqrt{6}} \Lambda + \frac{1}{\sqrt{2}} \Sigma^0& \Sigma^+ & p\\
\Sigma^-&\frac{1}{\sqrt{6}} \Lambda- \frac{1}{\sqrt{2}} \Sigma^0 &n\\
\Xi^- &\Xi^0 & -\sqrt{\frac{2}{3}} \Lambda 
\end{array}\right).
\end{eqnarray}
The constants $F^\prime = 0.46$ and $D^\prime = 0.8$, in Eq.~(\ref{LPB}), reproduce the axial coupling constant of the nucleon:  $F^\prime + D^\prime \simeq  g_A = 1.26$.

Using this Lagrangian, we compute the following amplitudes for the contact interaction and for diagrams involving the exchange of a $1/2^+$ octet baryon in the $s$- and $u$-channel, which are in agreement with other works~\cite{Oller:2000fj,Guo,Borasoy:2005ie,Ramos:2016odk},
\begin{align}
V_{\rm cont}& (i\to j) = -\frac{1}{4f_P^2} \sqrt{\frac{M_i + E_i}{2M_i}}\sqrt{\frac{M_j + E_j}{2M_j}}  \mathcal A_{ij} \Biggl[ \left(2\sqrt{s}-M_i -M_j \right) +\biggl(2\sqrt{s}+M_i+M_j\biggr)\Biggr.\nonumber\\
&\times\Biggl.\left(\frac{\vec p_i \cdot \vec{p}_j + i ~\chi^\dagger_j ~\left(\vec p_j \times \vec p_i\right) \cdot  \vec \sigma~ \chi_i}{(M_i + E_i)(M_j + E_j)}\right)\Biggr],\label{Eq:WT}
\end{align}
\begin{align}
V_s& (i\to j)  = \frac{1}{2f_P^2} \sqrt{\frac{M_i + E_i}{2M_i}}\sqrt{\frac{M_j + E_j}{2M_j}} \sum_k \frac{\mathcal B_{ij}^k}{s - M_{k}^2} \Biggl[ \left(\sqrt{s}-M_i \right)\left(\sqrt{s}-M_j \right)\left(\sqrt{s}-M_k \right)\Biggr.\nonumber\\
&+\Biggl.\left(\frac{\vec p_i \cdot \vec{p}_j + i ~\chi^\dagger_j ~\left(\vec p_j \times \vec p_i\right) \cdot  \vec \sigma~ \chi_i}{(M_i + E_i)(M_j + E_j)}\right)\left(\sqrt{s}+M_i \right)\left(\sqrt{s}+M_j \right)\left(\sqrt{s}+M_k \right)\Biggr],\label{Eq:s}\\\nonumber\\
V_u& (i\to j) = -\frac{1}{2f_P^2} \sqrt{\frac{M_i + E_i}{2M_i}}\sqrt{\frac{M_j + E_j}{2M_j}} \sum_k \frac{\mathcal C_{ij}^k}{u - M_{k}^2} \Biggl[ u\left(\sqrt{s} + M_{k}\right) +\sqrt{s}\biggl(M_j\left[M_i+M_k\right] \biggr. \Biggr.\nonumber\\
&+M_i M_k\biggr)  - M_j\left(M_i + M_k\right)\left(M_i+M_j\right) - M_i^2 M_k + \left(\frac{\vec p_i \cdot \vec{p}_j + i ~\chi^\dagger_j ~\left(\vec p_j \times \vec p_i\right) \cdot  \vec \sigma~ \chi_i}{(M_i + E_i)(M_j + E_j)}\right)\nonumber\\
&\times\Biggl.\biggl(u \left(\sqrt{s}-M_k\right) + \sqrt{s}\left(M_j\left[M_i+M_k\right] + M_i M_k\right) + M_j\left(M_i+M_j\right)\left(M_i+M_k\right)+M_i^2M_k\biggr)\Biggr].\label{Eq:u}
\end{align}
The summation in Eqs.~(\ref{Eq:s})~and~(\ref{Eq:u}) corresponds to summing the diagrams with different allowed octet baryons exchanged in the $s$-, $u$-channel, respectively, for a given process $i \to j$, with $i$ ($j$) [here, and in Eqs.~({\ref{Eq:WT}),~(\ref{Eq:s}),~and~\ref{Eq:u})] representing the initial (final) state. In these equations, $M_l$ ($E_l$) denotes the mass (energy) of the baryon in the initial/final/intermediate state, represented by a subindex $l = i/j/k$, respectively, $\vec p_l$ represents the center of mass momentum in the $l$th channel and $\mathcal A_{ij}$, $\mathcal B_{ij}$, $\mathcal C_{ij}$ are isospin coefficients for different processes. The coefficients $\mathcal B_{ij}$, $\mathcal C_{ij}$, for isospin 0 and 1, are listed in Tables.~\ref{iso0s1hdirect}, \ref{iso1s1hdirect}, \ref{iso0s1hcross},~and~\ref{iso1s1hcross} in the Appendix, where we also give the amplitudes in Eqs.~(\ref{Eq:WT}), (\ref{Eq:s}), and (\ref{Eq:u}) projected on $s$-wave. We refer the reader to Ref.~\cite{osetramos} for the constants, $\mathcal A_{ij}$, related to the contact interactions. It must be added here that we consider an octet baryon exchange in the $s$- and $u$-channel, thus, the  $1/2^-$ states eventually found in the complex plane can be interpreted as those arising from the dynamics in the system.

For the vector-baryon amplitudes, we follow the previous work~\cite{vbvb}, where the problem was studied in detail, using a Lagrangian based on hidden local symmetry, and it was found that $s$-, $t$-, and $u$-channel diagrams and a contact interaction arising from two vector field terms give comparable contributions, and must all be considered. We take the following Lagrangian from Ref.~\cite{vbvb}:  
\begin{eqnarray} \label{vbb}
&\mathcal{L}_{\textrm VB}& = -g \Biggl\{ \langle \bar{B} \gamma_\mu \left[ V_8^\mu, B \right] \rangle\!+\! \langle \bar{B} \gamma_\mu B \rangle  \langle  V_8^\mu \rangle  
\Biggr. + \frac{1}{4 M}\! \left( F \langle \bar{B} \sigma_{\mu\nu} \left[ V_8^{\mu\nu}, B \right] \rangle \! +\! D \langle \bar{B} \sigma_{\mu\nu} \left\{V_8^{\mu\nu}, B \right\} \rangle\right)\\\nonumber
&& +  \Biggl.  \langle \bar{B} \gamma_\mu B \rangle  \langle  V_0^\mu \rangle  
+ \frac{ C_0}{4 M}  \langle \bar{B} \sigma_{\mu\nu}  V_0^{\mu\nu} B  \rangle  \Biggr\},
\end{eqnarray}
where the subscript $8$ ($0$) denotes the octet (singlet) part of the wave function of the vector meson (relevant in the case of $\omega$ and $\phi$), $V^{\mu\nu}$ represents the  tensor field of the vector mesons,
\begin{equation}
V^{\mu\nu} = \partial^{\mu} V^\nu - \partial^{\nu} V^\mu + ig \left[V^\mu, V^\nu \right], \label{tensor}
\end{equation}
 and $V^\mu$ is the SU(3) matrix for the (physical) vector mesons,
\begin{eqnarray}
V^\mu &=&\frac{1}{2}
\left( \begin{array}{ccc}
\rho^0 + \omega & \sqrt{2}\rho^+ & \sqrt{2}K^{*^+}\\
\sqrt{2}\rho^-& -\rho^0 + \omega & \sqrt{2}K^{*^0}\\
\sqrt{2}K^{*^-} &\sqrt{2}\bar{K}^{*^0} & \sqrt{2} \phi 
\end{array}\right)^\mu.
\end{eqnarray}
In Eq.(\ref{vbb}), the coupling $g$ is related to the vector meson decay constant, $f_v$ through the Kawarabayashi-Suzuki-Riazuddin-Fayyazuddin (KSRF) relation,
\begin{align}
g=\frac{m_v}{\sqrt{2}f_v}, \label{ksrf}
\end{align}
with $m_v$ being the mass of a given vector meson in the vertex and the constants $D$ = 2.4, $F$ = 0.82 and $C_0 = 3F - D$  correctly reproduce the anomalous magnetic 
couplings of the $\rho NN$, $\omega NN$ and $\phi NN$ vertices~\cite{jenkins,meissner,jido}. Together with Eq.~(\ref{vbb}), and the kinetic term
\begin{equation}
\mathcal{L}_{3V} \in - \frac{1}{2} \langle V^{\mu\nu} V_{\mu\nu} \rangle,
\end{equation}
it is possible to calculate the $s$-, $t$-, and $u$-channel amplitudes as well as the contact interaction by using $\left[V^\mu, V^\nu \right]$ for $V^{\mu\nu}$ in Eq.~(\ref{vbb}).
It was found in Ref.~\cite{vbvb} that this contact interaction, apart from giving contributions comparable to other amplitudes, is important to guarantee the  invariance of the Lagrangian under a gauge transformation.

Finally, the amplitudes for the transition between the pseudoscalar-baryon and the vector-baryon channels are deduced from the Lagrangian~\cite{pbvb}
\begin{eqnarray} \label{pbvbeq}
\mathcal{L}_{\rm PBVB} &=& \frac{-i g_{PBVB}}{2 f_v} \left( F^\prime \langle \bar{B} \gamma_\mu \gamma_5 \left[ \left[ P, V^\mu \right], B \right] \rangle \right. +
\left. D^\prime \langle \bar{B} \gamma_\mu \gamma_5 \left\{ \left[ P, V^\mu \right], B \right\}  \rangle \right),
\end{eqnarray}
which has been obtained by introducing the vector meson field as a gauge boson of the hidden local symmetry in the nonlinear sigma model. The procedure is, thus, like 
extending the Kroll-Ruderman term for the photoproduction of a pion, replacing, inspired by the vector meson dominance, the photon by the vector meson~\cite{pbvb}. The constants,  $F^\prime$ and $D^\prime$ are the same as those defined
for Eq.~(\ref{LPB}).

The formalism has been applied to study meson-baryon systems with various quantum numbers in Refs.~\cite{Khemchandani:2012ur,nstars,Khemchandani:2016ftn} and, in fact, different vector-baryon amplitudes as well as those for the transition between pseudoscalar- and vector-baryon channels are taken from Ref.~\cite{Khemchandani:2012ur} for the present work. Though, it must be mentioned that the formalism in the present work is more elaborate, as compared to our previous works, since we include $s$- and $u$-channel octet baryon exchange diagrams for pseudoscalar-baryon interactions here. The contributions from these diagrams have been found to play an important role in the formation of isospin one resonances near 1400 MeV \cite{Oller:2000fj,Guo} and it is the  purpose of the present work to constrain our amplitudes to reproduce the experimental data in the low-energy region and investigate the formation of isospin 1 states around 1400 MeV.

To proceed further, we unitarize the tree-level meson-baryon amplitudes calculated from the Lagrangians introduced. The resulting expressions \cite{Oller:2000fj,Oller:1998zr} are the same as obtained by factorizing on-shell the potential in a Bethe-Salpeter equation \cite{osetramos,Oller:1997ti}. Though used extensively, a few words to motivate the on-shell factorized form are in order here. Such a method is inspired by the fact that for a contact$-$like interaction potential  $V$ , when projected in $s$-wave and iterated in the equation, produces an off-shell dependence which leads to tadpole-type loop integrals whose contributions can be absorbed in the parameters, like the pion decay constant, appearing in the kernel (see, for example, Ref.~\cite{Oller:1997ti}). Thus, one could work with an $s$-wave projected potential $V$ in which the original pion decay constant, which could be considered as a kind of bare decay constant, is substituted by a dressed one and eliminate the associated tad-pole Feynman diagrams related to the off-shell part of the kernel.  Another  motivation for the on-shell approximation comes from the two-body unitarity in  coupled channels and implementation of a dispersion relation for the imaginary part of the inverse of the $T$-matrix considering the physical (or unitarity) cut (see, for example, Ref.~\cite{Oller:1998zr}). In both cases, a divergent loop function of two hadrons appears and needs to be regularized. The  method differs from the field-theoretical standard procedure of canceling such divergences by adding counter-terms in the Lagrangian used to determine the kernel $V,$ a fact which gets reflected in the regularized loop function through the appearance of the unknown subtraction constants, which need to be fixed, for example, by fitting the data. By fixing them, one is somehow generating the counter-terms in the on-shell factorization scheme, since such a subtraction constant can be somehow reabsorbed in the kernel $V$, when iterating it in the Bethe-Salpeter equation, producing a new kernel. This procedure is explained in detail in  chapter 7 of Ref.~\cite{libro_oller}. When the potential $V$ also contains crossed-channel cuts this unitarization procedure can be applied by matching algebraically (not numerically) the unitarized result with the perturbative one order by order, as explained, e.g., in Refs.~\cite{Oller:2000fj,Oller:1999me} or in  chapter 10 of Ref.~\cite{libro_oller}. This is the method that we are using here where $V$ also contains the $u$-channel exchange of the lightest $1/2^+$ octet of baryons. 
 Either way, the so called on-shell factorization has been remarkably successful in understanding, reproducing and predicting the properties of many of the resonances observed in nature, like $\sigma$(600), $f_0$(980), $a_0$(980), $\Lambda$(1405), etc., proving its reliability in a study like the one at hand.

With the lowest order amplitudes discussed in this section we solve the Bethe-Salpeter equation in its on-shell factorized form and make a $\chi^2$-fit to the data. The parameters of the fit are:
\begin{enumerate} \item{The subtraction constants required to calculate the loop integrals with the dimensional-regularization method
 \begin{eqnarray}
G (\sqrt{s}) &=& i 2 M \int \frac{d^4q}{2\pi^4} \frac{1}{(\tilde P - q)^2 - M^2 + i\epsilon} \frac{1}{q^2 - m^2 + i\epsilon} \label{loop} \\\nonumber
&=&\frac{2M}{16\pi^2} \Biggl\{ a (\mu) + \ln \frac{M^2}{\mu^2} + \frac{m^2-M^2+s}{2s}\ln \frac{m^2}{M^2} \Biggr.\\\nonumber
&+& \frac{\tilde{q}}{\sqrt{s}} \Bigl[ \ln\left(s- \left( M^2 - m^2 \right) + 2\tilde{q}\sqrt{s}\right) \Bigr.
+  \ln\left(s+ \left( M^2 - m^2 \right) + 2\tilde{q}\sqrt{s}\right) \\\nonumber
&-& \ln\left(-s +\left( M^2 - m^2 \right) + 2\tilde{q}\sqrt{s}\right) 
- \Biggl. \Bigl. \ln\left(s- \left( M^2 - m^2 \right) + 2\tilde{q}\sqrt{s}\right) \Bigr] \Biggr\},
 \end{eqnarray}
 where $\tilde P$ is the total four-momentum, $M$ ($m$) is the mass of the propagating baryon (meson), and $\tilde{q} = \lambda^{1/2} (s, M^2,m^2) /2\sqrt{s}$, $a(\mu)$ is the subtraction constant at a regularization scale $\mu = 630$~MeV. 
In line with the discussions on the on-shell factorization form of the Bethe-Salpeter equation, the implementation of  coupled channel unitarity relates the imaginary part of the inverse of the  $T$-matrix to the  phase space for the corresponding elastic transition. In this way, when implementing a dispersion relation for the inverse of the $T$-matrix with a constant subtraction, it is expected to have one subtraction constant for each elastic transition, i.e, one subtraction constant for each coupled channel. Using isospin average masses for members of the same isospin multiplet, we would have a subtraction constant for each coupled channel in the isospin base.

Since a fit is made to both isospin 0 and 1 amplitudes, we have 14 subtraction constants as parameters, corresponding to the channels: $\bar KN$, $K \Xi$, $\pi \Sigma$, $\eta \Lambda$, $\pi \Lambda$, $\eta \Sigma$, $\bar K^*N$, $K^* \Xi$, $\rho \Sigma$, $\omega \Lambda$, $\phi \Lambda$, $\rho \Lambda$, $\omega \Sigma$, $\phi \Sigma$.}
 \item{The decay constants of the mesons. In general, different mesons have different decay constants. 
We could fix the pseudoscalar and vector-meson decay constants to their physical values. However, it is quite common in this kind of calculations to use a unique value for the three pseudoscalar decay constants, which typically corresponds to the value of 93 MeV, which is the pion decay constant (see, for example, Ref.~\cite{Oller:1997ti}), or an average of their physical values for all of them (see, for example, Refs.~\cite{osetramos,Jido:2003cb}). The same thing can be said for the vector decay constants. The motivation being that a change in the value of the potential obtained at  ${\cal O}(p)$ by modifying the decay constant used, can be reabsorbed in the value of the subtraction constant used to regularize the loop, producing mild modifications on its value.  In the present work, where we we unitarize the tree -level amplitudes at ${\cal O}(p)$ and in which the subtraction constants are obtained by fitting the data, we have continued adopting the widely used strategy of using an average value for the pseudoscalar decay constants and an average value for the vector decay constants. We allow them to vary mildly from their physical values, so that such an option could correspond to higher order corrections of the ${\cal O}(p)$ $T$ matrix. 
 
Besides, not all decay constants are well known, as is the case, for instance, of the $K^*$-meson (see Ref.~\cite{Chang:2018aut} for one of the latest calculations from lattice). Even in the case of pseudoscalars, the extraction of a precise value of the $\eta$-meson decay constant still seems to be under investigation~\cite{pdg,Feldmann:1999uf}.

 Thus, the average value for the decay constant of the pseudoscalars, $f_P$, to be used in Eqs.~(\ref{Eq:WT_swave}),~(\ref{Eq:s_swave}),~(\ref{Eq:u_swave}), and another one for the vectors, $f_v$, to be used in vector-baryon amplitudes,  account for two additional parameters in the fit.}
 \item{Finally, the coupling at the pseudoscalar-baryon--vector-baryon vertex, $g_{PBVB}$ in Eq.~(\ref{pbvbeq}), is treated as a parameter to be fitted,  whose value can be approximately estimated using Eq.~(\ref{ksrf}). One gets $g_{PBVB}\sim$ 3.5 by taking an average value for  $m_v \sim$ 850 MeV, $f_v \sim$ 170 MeV. However, this value could be smaller if hadronic structure is taken into account by using a form factor. Note that  if the pion decay constant $\sim$ 93 MeV is used, instead of the vector decay constant, in Eq.~(\ref{ksrf}), then $g_{PBVB} \sim6$ (as in Refs.~\cite{pbvb,Khemchandani:2012ur,nstars}). We, thus, allow  $g_{PBVB}$ to vary between 1 and 6 in the fitting procedure. }
\end{enumerate}

The experimental data considered for the fit are: 
\begin{enumerate}
\item{The total cross sections of the processes: $K^- p \to K^- p,~\bar K^0 n,~ \eta \Lambda,~\pi^0 \Lambda~\pi^0 \Sigma^0,~ \pi^\pm \Sigma^\mp$, from the respective thresholds to about 30-50 MeV above the threshold \cite{Starostin:2001zz,Humphrey:1962zz,Ciborowski:1982et,Evans:1983hz,Kim:1967zze,Sakitt:1965kh,Kittel:1966zz}. }
\item{The energy level shift and width of the $1s$ state of the kaonic hydrogen measured by the SIDDHARTA collaboration~\cite{Bazzi:2011zj}: $\Delta E = 283 \pm 36 \pm 6$ eV and  $\Gamma = 549 \pm 89 \pm 22$~eV. We use the relation between the energy shift and width of the $1s$ state of the kaonic hydrogen and the $K^-p$ scattering length, as obtained in Ref.~\cite{Meissner:2004jr}
\begin{equation}
\Delta E  - i \frac{\Gamma}{2} = - 2 \alpha^3 \mu^2 a_{K^-p}\left[1 + 2 \alpha \mu (1 - ln \alpha) a_{K^-p}\right] ,\label{1s_energy}
\end{equation}
where
\begin{equation}
a_{K^-p} = - \frac{t_{K^-p}}{4 \pi \sqrt{s_{th}}} M_p,\label{aeq}
\end{equation}
with $M_p$ being the proton mass and $\sqrt{s_{th}}$ denoting the $K^-p$ threshold energy.   }
\item{The following ratios of the  cross section  at the threshold, taken from Refs.~\cite{Tovee:1971ga,Nowak:1978au},
\begin{align}\nonumber
\gamma&=\frac{\sigma(K^- p\to \pi^+\Sigma^-)}{\sigma(K^- p\to \pi^-\Sigma^+)} = 2.36 \pm 0.12,\\
R_c&=\frac{\sigma(K^- p\to\text{charged particles})}{\sigma(K^- p\to \text{all})} = 0.664\pm0.033,\label{ratios}\\\nonumber
~R_n&=\frac{\sigma(K^-p\to\pi^0\Lambda)}{\sigma(K^- p\to\text{all neutral states})}=0.189\pm0.015,
\end{align}
where,  following Ref.~\cite{Guo},
a conservative 5 $\%$ relative error bar is assigned to the value of $\gamma$, $R_c$ to include the different experimental measurements.}
\end{enumerate}

\section{Results and discussions}\label{results} 
To fit the data, the $\chi^2$ per degree of freedom, $\chi^2_\text{d.o.f}$, is calculated as~\cite{Borasoy:2005ie,Guo,GarciaRecio:2002td,Ikeda:2012au,Mai:2012dt}, 
\begin{align}
\chi^2_\text{d.o.f}=\frac{\sum\limits_{k=1}^N n_k}{N (\sum\limits_{k=1}^N n_k-n_p)}\sum\limits_{k=1}^N \frac{\chi^2_k}{n_k},
\end{align}
where $N$ is the number of different data sets, $n_k$ represents the number of data points in the $k$th data set, $n_p$ is the number of free parameters, and the $\chi^2$ for the $k$th data set is obtained as
\begin{align}
\chi^2_k=\sum\limits_{i=1}^{n_k}\frac{(y^{\text{th}}_{k;i}-y^{\text{exp}}_{k;i})^2}{\sigma^2_{k;i}},
\end{align}
with $y^{\text{exp}}_{k;i}$ ($y^{\text{th}}_{k;i}$) representing the $i$th experimental (theoretical) point of the $k$th data set and $\sigma^2_{k;i}$ the standard deviation associated with it. In this context, we should mention that the values of  $\Delta E$ and $\Gamma$ from the SIDDHARTA collaboration are considered as two data points of the same data set.

In the fitting procedure, we find that two types of solutions exist,  which correspond to $\chi^2_\text{d.o.f} \sim 1$. A $\chi^2_\text{d.o.f}$  value of the order of 1 is the expected value for such a quantity when the number of degrees of freedom is large, with the $\chi^2$ having a standard deviation of one.  Large deviations from 1 for $\chi^2_\text{d.o.f}$  would imply, thus, that the fit found to the data could be categorized as a bad fit. The parameter sets related to the two solutions, which we label  as Fit~I~and~II, are given in Table~\ref{par},  
together with the associated error bars. The central value and the associated error correspond to the mean value and the standard deviation, respectively, obtained for each parameter. The errors are estimated by admitting  solutions satisfying the condition 
\begin{align}
\chi^2\leqslant\chi^2_0+\sqrt{2\chi^2_0},\label{criteria}
\end{align} 
where $\chi^2_0$ is the minimum $\chi^2$ value obtained, as in Refs.~\cite{Albaladejo:2008qa,Etkin:1981sg}.  Equation~(\ref{criteria}),  obtained in Ref.~\cite{Etkin:1981sg}, is based on the fact that in the limit of large number of degrees of freedom, ($\chi^2-$n.d.o.f)/($\sqrt{2~\text{n.d.o.f}}$) is normally distributed with the mean value 0 and standard deviation 1. Thus, producing random numbers, within the error bars, for the parameters obtained from the best fit and considering all the new fits  satisfying Eq.~(\ref{criteria}) implies estimating the parameter to a confidence level of 1 standard deviation and including, at the same time, the correlated errors of all the free parameters.

Besides the above discussions, we must add that the biggest contribution to the $\chi^2_\text{d.o.f}$ comes from the cross section data for the different $K^- p$ processes mentioned in the previous section. Thus, when minimizing the $\chi^2_\text{d.o.f}$ it is possible to find solutions with $\chi^2_\text{d.o.f} \sim 1$, but the values obtained for the ratios of Eq.~(\ref{ratios}) and/or the SIDDHARTA data lay outside the error bars related to the respective experimental data, making the results from the fit incompatible with these data, in spite of fitting well the cross sections data on  the $K^- p$ processes. Such fits have been discarded.
{\squeezetable \begin{table}[h!]
\caption{Values of the parameters obtained by constraining the model amplitudes to reproduce experimental data  (mentioned in Sec.~\ref{sec:formalism}). 
Here, $a_i$ represents the subtraction constant for the channel $i$ in the isospin base, $f_P$ ($f_v$) is an average value for the decay constants of the pseudoscalar (vector) mesons, and $g_{PBVB}$ is the coupling appearing in the $PB\leftrightarrow VB$ vertices [see Eq.~(\ref{pbvbeq})]. The values of the minimum $\chi^2_\text{d.o.f}$ are 0.89 for Fit I and 0.91 for Fit II.
}\label{par}
\vspace{0.2cm}
\centering
\begin{tabular}{ccc}
\hline
Parameters&Fit I& Fit II\\
\hline\hline
$a_{\bar K N}$& $-2.00\pm0.06$&$-2.12\pm0.10$\\
$a_{K \Xi}$& $-2.43\pm0.04$&$-2.43\pm0.06$\\
$a_{\pi\Sigma}$&$-1.09\pm0.07$&$-1.18\pm0.12$ \\
$a_{\eta \Lambda}$& $-1.25\pm0.03$&$-1.27\pm0.09$\\
$a_{\pi \Lambda}$& $-0.84\pm0.26$&$-1.69\pm0.31$\\
$a_{\eta\Sigma}$&$-3.62\pm0.44$&$-1.97\pm0.12$ \\\hline
\end{tabular} 
\begin{tabular}{ccc}
\hline
Parameters&Fit I&Fit II\\
\hline\hline
$a_{\bar K^* N}$& $-4.34\pm0.08$&$-4.39\pm0.09$\\
$a_{K^* \Xi}$& $-3.86\pm0.03$&$-3.33\pm0.06$\\
$a_{\rho\Sigma}$& $1.17\pm1.29$&$-2.36\pm0.07$\\
$a_{\omega \Lambda}$& $-6.50\pm0.70$&$-3.86\pm2.09$\\
$a_{\phi\Lambda}$& $-6.83\pm0.60$&$-5.22\pm1.13$\\
$a_{\rho\Lambda}$&$-0.77\pm0.20$&$-0.49\pm0.47$ \\\hline
\end{tabular}\vspace{0.4cm}
\begin{tabular}{ccc}
\hline
Parameters&Fit~I&Fit~II\\
\hline\hline
$a_{\omega\Sigma}$& $-3.55\pm1.58$&$-3.65\pm1.34$\\
$a_{\phi\Sigma}$&$-4.67\pm0.29$&$-2.51\pm0.39$ \\
$f_P~(\text{MeV})$& $94.62\pm1.46$&$97.24\pm1.56$\\
$f_v$~(\text{MeV})&$138.12\pm1.54$&$113.46\pm5.21$ \\
$g_{PBVB}$& $2.19\pm0.09$&$1.81\pm0.07$\\
&&\\
\hline
\end{tabular}
\end{table}}

In Fig.~\ref{fit1} 
\begin{figure}[h!]
\includegraphics[width=0.35\textwidth]{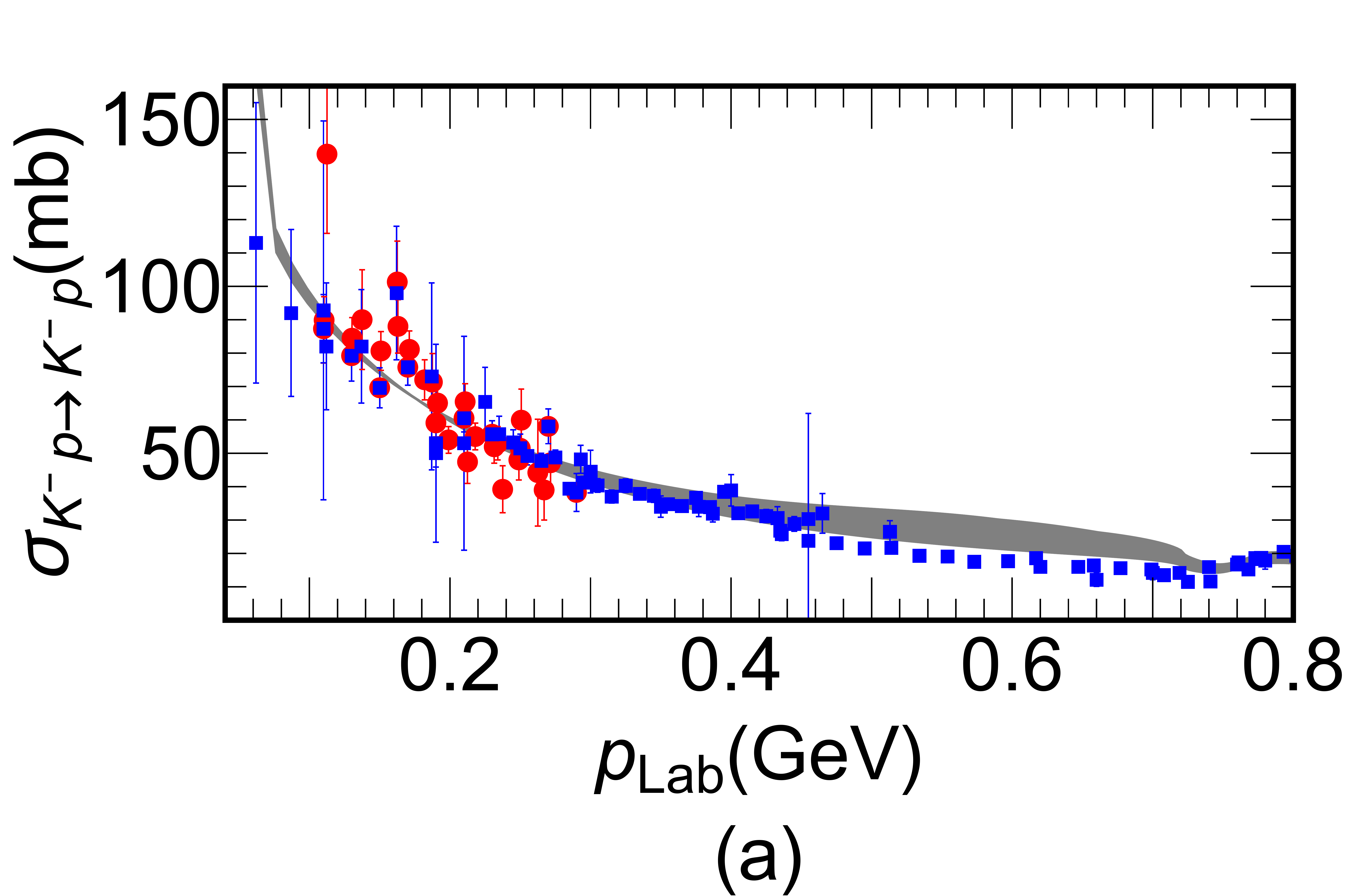}
\includegraphics[width=0.35\textwidth]{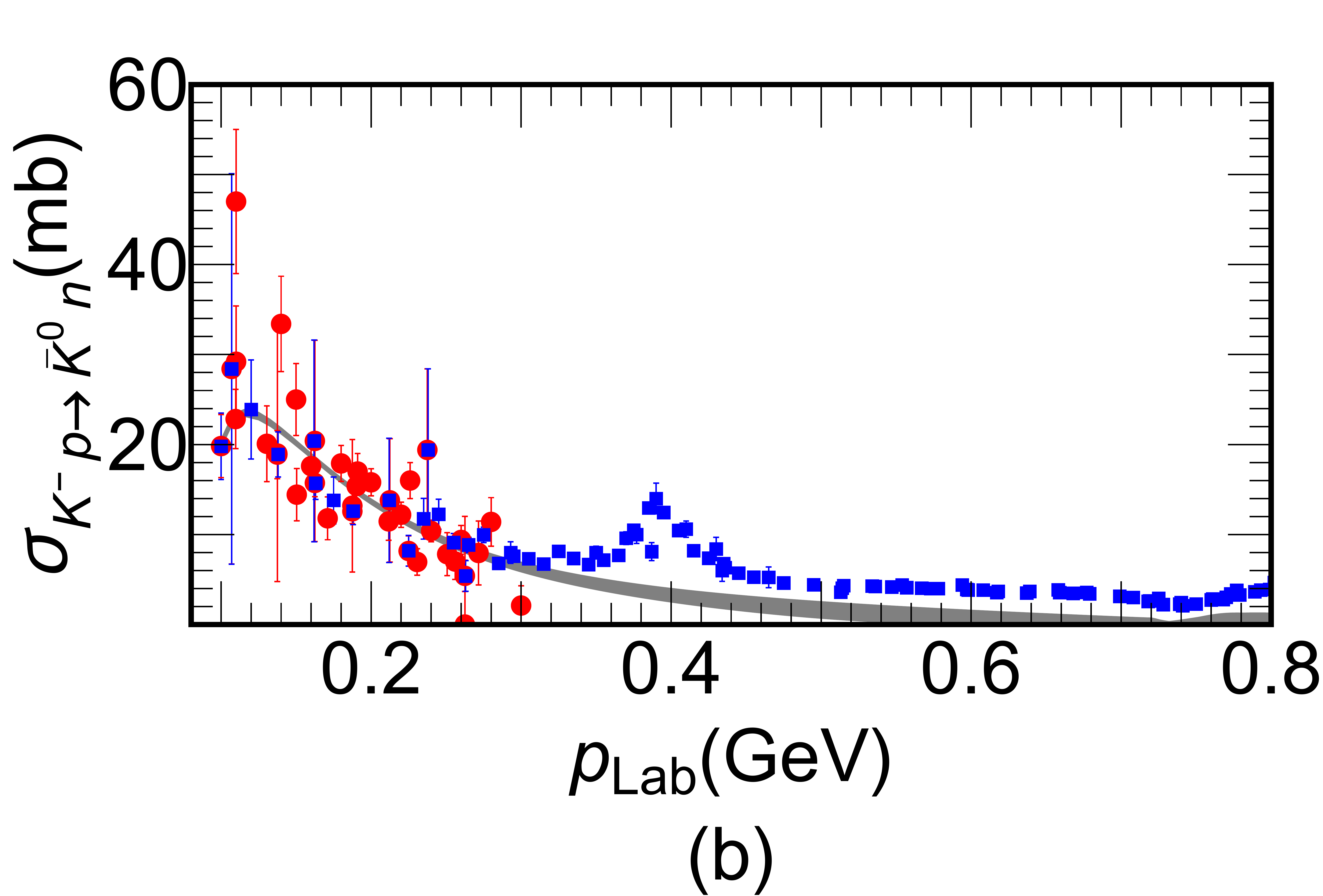}
\includegraphics[width=0.35\textwidth]{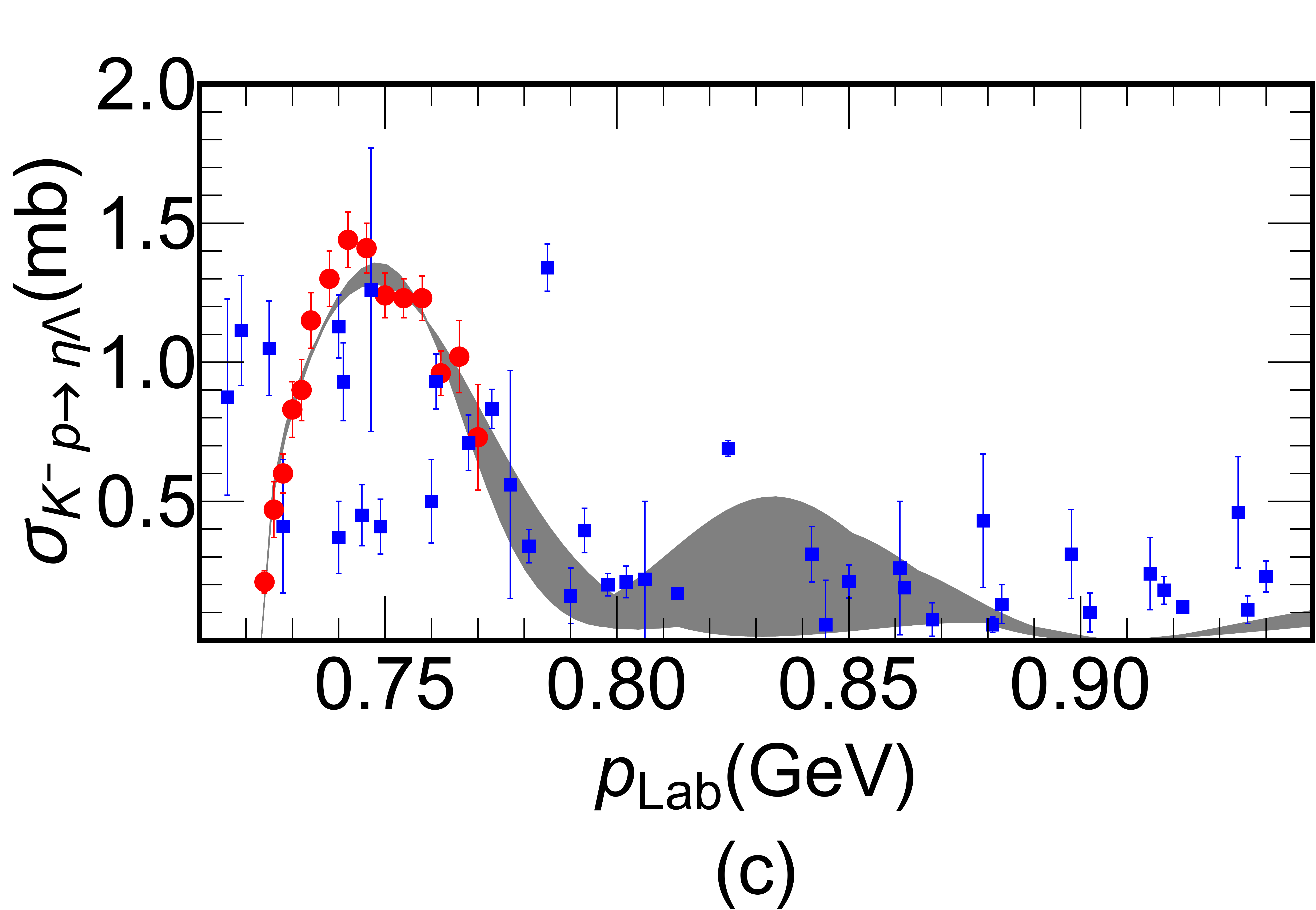}
\includegraphics[width=0.35\textwidth]{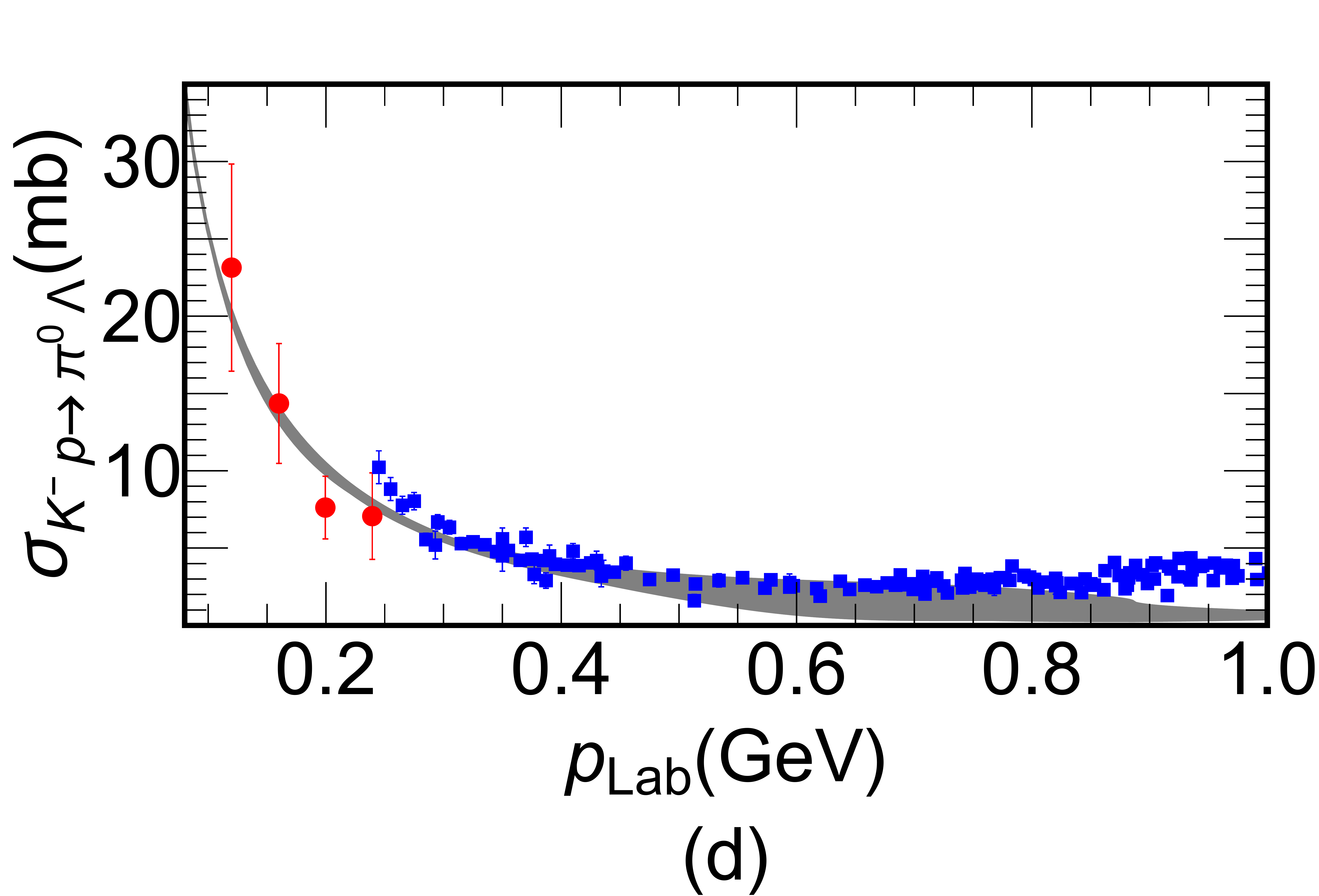}
\includegraphics[width=0.35\textwidth]{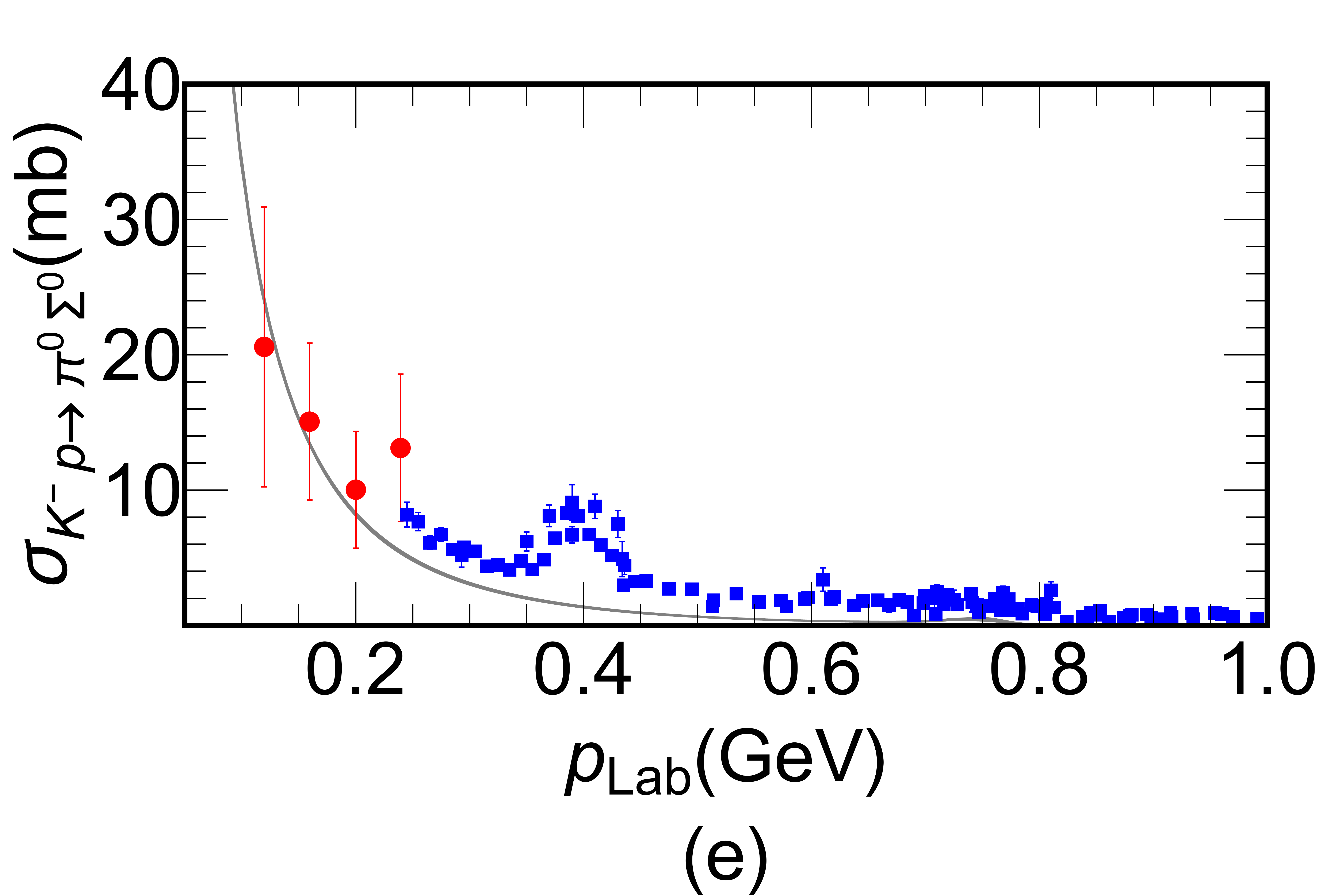}
\includegraphics[width=0.35\textwidth]{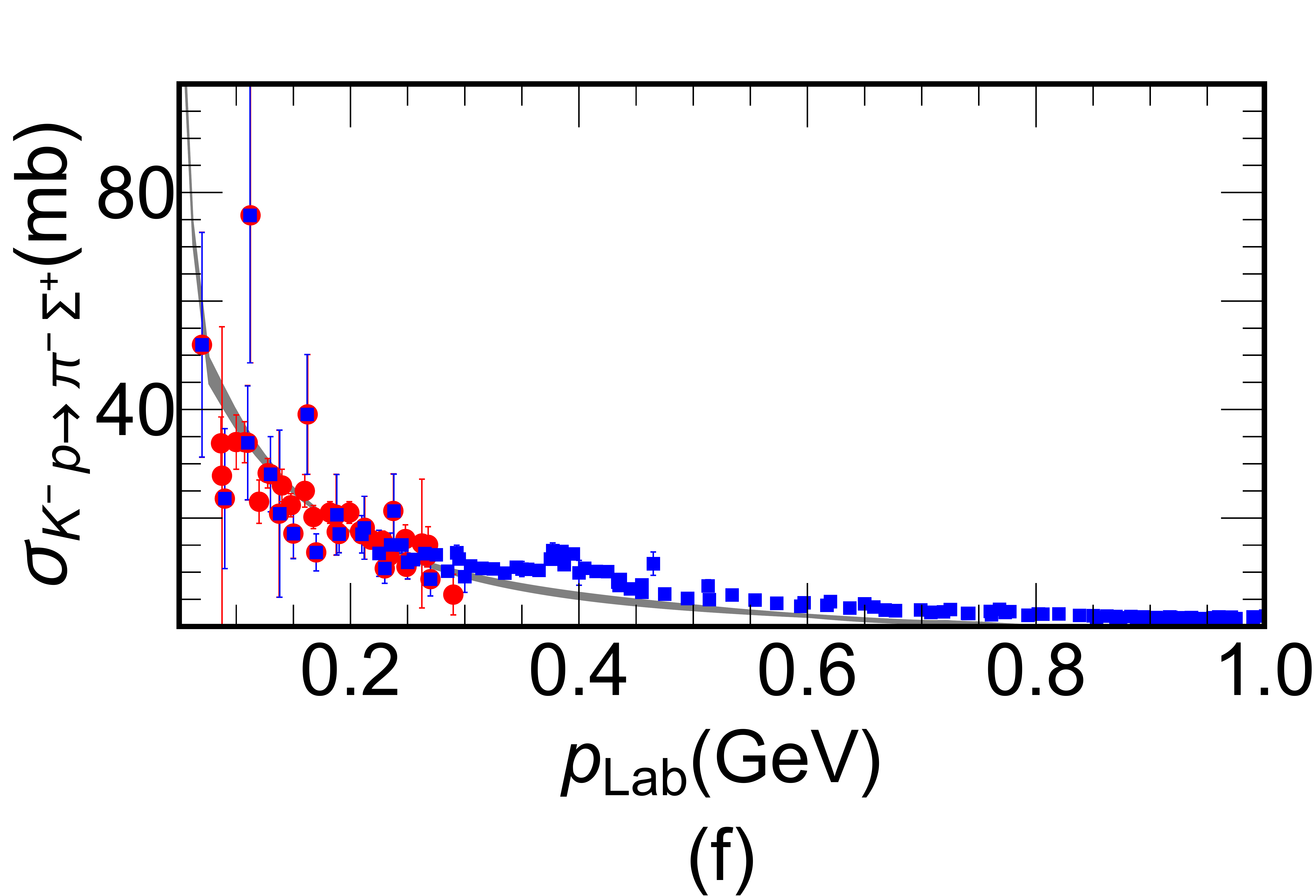}
\includegraphics[width=0.35\textwidth]{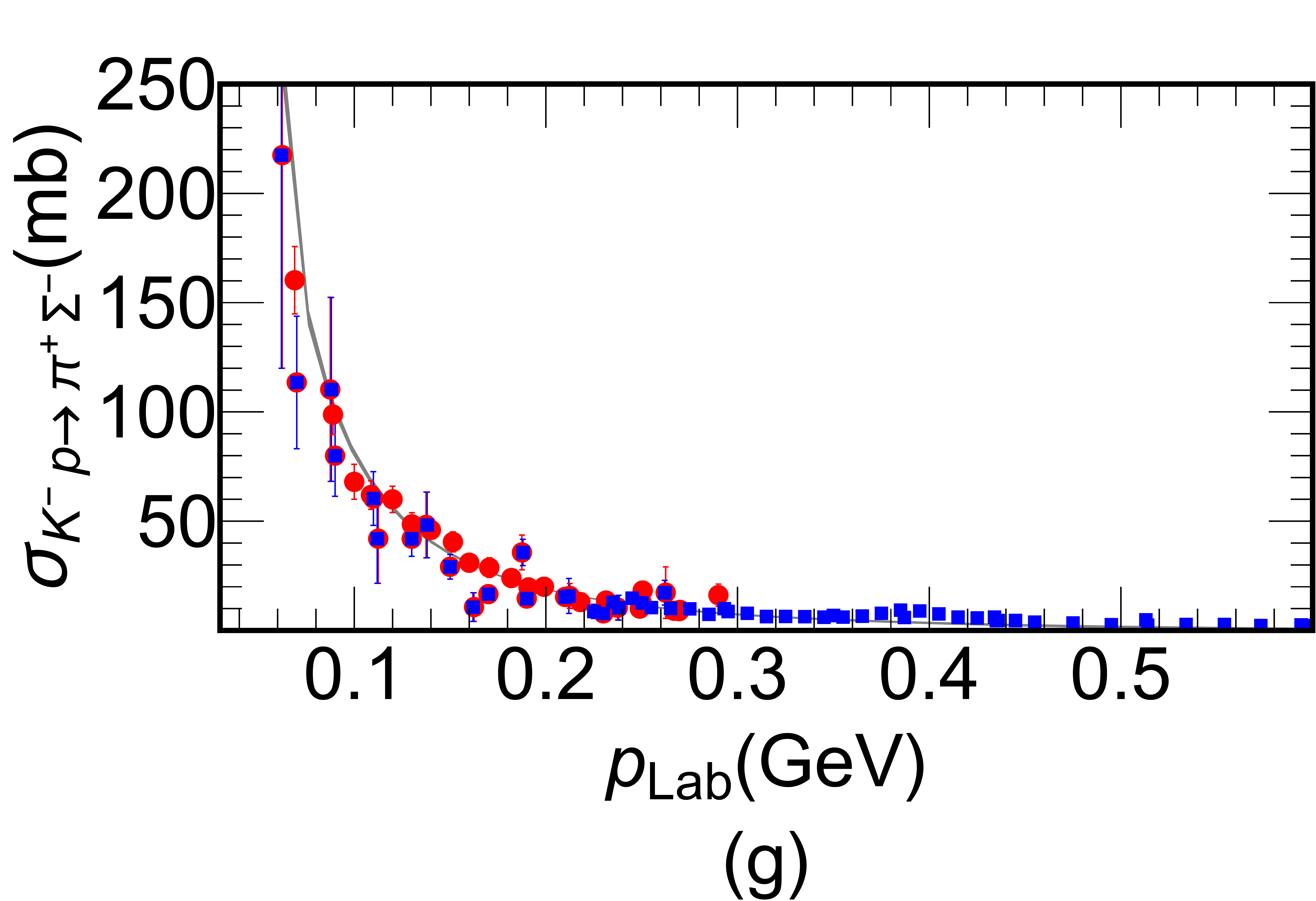}
\caption{Cross sections of different processes studied in our work. The shaded region represents the results found with the parameters listed under the label Fit I, in Table~\ref{par}.  Data shown as (red) filled circles (taken from Refs.~\cite{Starostin:2001zz,Humphrey:1962zz,Ciborowski:1982et,Evans:1983hz,Kim:1967zze,Sakitt:1965kh,Kittel:1966zz}) were used in the $\chi^2$ fitting procedure explained in the text.}\label{fit1}
\end{figure}
we show the cross sections of the different processes, as obtained by the parameter set labeled as  Fit I. 
The shaded bands in the panels correspond to the results obtained by using the criteria given in Eq.~(\ref{criteria}).  The data considered in the fit are shown as (red) filled circles in Fig.~\ref{fit1}. These data are taken from Refs.~\cite{Starostin:2001zz,Humphrey:1962zz,Ciborowski:1982et,Evans:1983hz,Kim:1967zze,Sakitt:1965kh,Kittel:1966zz}, and are the same as those considered in Ref.~\cite{Guo}. We have included more data points from Ref.~\cite{landolt} and which are shown as (blue) filled squares in Fig.~\ref{fit1}, going to about 100-200 MeV above the threshold for these reactions. It can be seen that the results stay close to the data points at higher energies too, even though the  data  at these energies were not used in the fit.  At energies farther from the reaction threshold, the cross sections are expected to get contributions from interactions in higher partial waves, and, thus, the $s$-wave amplitudes, which are the ones we calculate, are not expected to be sufficient to describe data at such energies. For a better description of the data we need to include some well-known resonances in the formalism, such as $\Lambda(1520) (3/2^-)$, $\Lambda(1600) (1/2^+)$, $\Sigma(1620) (1/2^+)$, which are related to $p$-, $d$-wave pseudoscalar-baryon interactions. Such states can be taken into account by including channels, like, meson--decuplet-baryon \cite{Roca:2006sz}, two meson-one baryon~\cite{MartinezTorres:2007sr}, etc. Such extensions of our work can be done in future. Still it is reassuring to see that the cross sections obtained at higher energies do not differ much from the experimental data.  It is worth mentioning that the coupling to vector-baryon channels is useful in improving this agreement, at energies away from the threshold. Although, the presence of the vector-baryon coupling is more significant in the case where the reaction threshold is higher (closer to the VB thresholds). Such is the case of the process $K^-p \to \eta \Lambda$, whose threshold is about 140 MeV away from the $\bar K^* N$, keeping in mind the finite width of  $K^*$. The finite widths of the vector mesons are taken into account in the formalism by folding the relevant loop function over the variable mass range of the vector mesons as \cite{ramosvb,vbvb} 
 \begin{eqnarray}
\tilde G_j (\sqrt{s}) = \dfrac{1}{N_j} \int\limits_{(m_j-2\Gamma_j)^2}^{(m_j +2\Gamma_j)^2} d\tilde{m}^2 \left(\dfrac{-1 }{\pi} \right) G_j(\sqrt{s})~
{\textrm Im} \left\{\dfrac{1}{\tilde{m}^2 - m_j^2 +  i m_j\Gamma(\tilde{m})} \right\},\label{Gconv}
\end{eqnarray}
where the subscript $j$ refers to the $j$th meson-baryon channel in the loop, $m_j$ ($\Gamma_j$) is the central mass (width) of the meson in the loop, $G_j(\sqrt{s})$ is calculated using Eq.~(\ref{loop}) and 
\begin{eqnarray}
N_j = \int\limits_{(m_j-2\Gamma_j)^2}^{(m_j+2\Gamma_j)^2}  d\tilde{m}^2 \left( - \dfrac{1}{\pi}\right) {\rm Im} \left\{\dfrac{1}{\tilde{m}^2 - m_j^2 + i m_j \Gamma(\tilde{m})}\right\}.\label{loopnorm}
\end{eqnarray}
The variable width in Eqs.~(\ref{Gconv})~and~(\ref{loopnorm}) for the $j$th meson decaying to mesons $a$ and $b$ is calculated as
\begin{equation}\nonumber
\Gamma(\tilde{m}) = \Gamma_{j} \left(\dfrac{m_j^2}{\tilde{m}^2}\right) \left(\dfrac{\lambda^{1/2}(\tilde{m}^2, m_a^2, m_b^{2})/2\tilde{m}} {\lambda^{1/2}(m_j^2, m_a^2, m_b^{2})/2m_j} \right)^3 \theta\left( \tilde{m} - m_a - m_b\right).
\end{equation}

In Fig.~\ref{Xn_compare},
\begin{figure}[h!]
\includegraphics[width=0.48\textwidth]{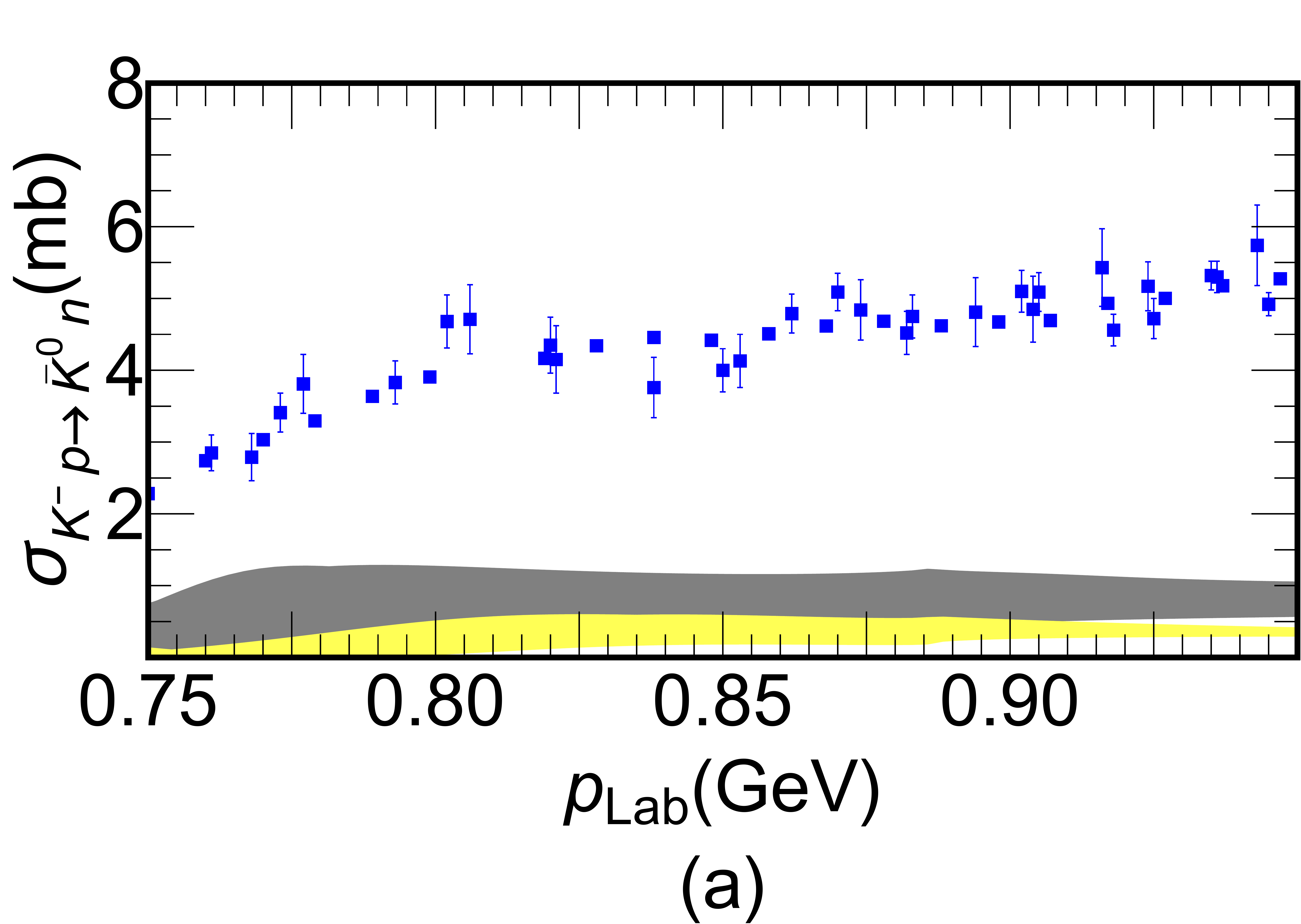}
\includegraphics[width=0.48\textwidth]{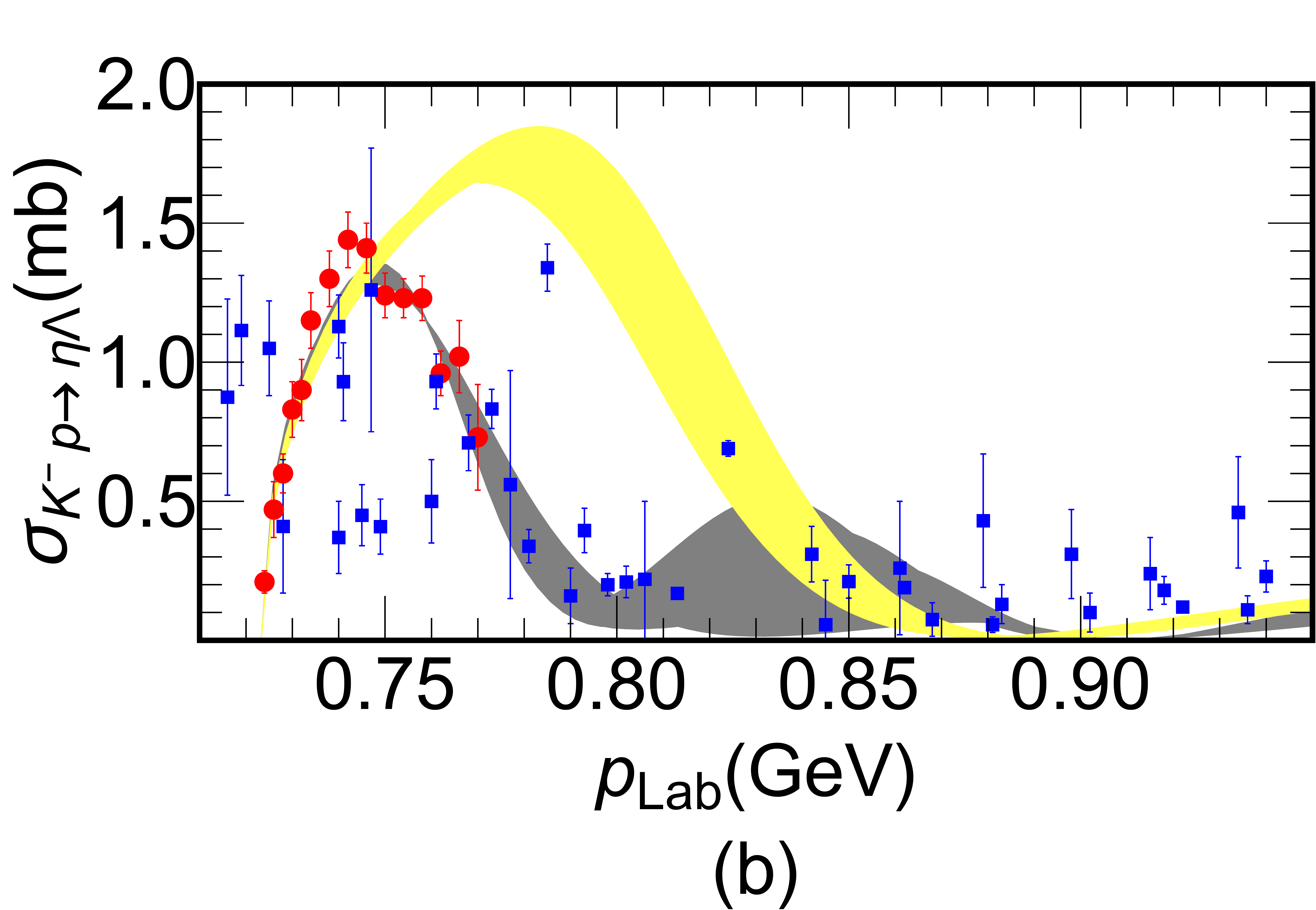}
\caption{A comparison of the cross sections obtained with (dark shaded) and without (light shaded) the coupling of pseudoscalar-baryon channels to vector-baryon channels.}\label{Xn_compare}
\end{figure}
we show the cross sections of the processes $K^-p \to \eta \Lambda$ and $K^-p \to \bar K^0 n$ obtained by decoupling  PB and VB channels, around the energy region where an influence of VB channels can be expected ($\sim$ 30-150 MeV below the lowest VB threshold). As can be seen, the coupling to the VB channels plays a more important role in the case of the process with a higher threshold.  

Before discussing the results  found, within the Fit~I,  for the energy shift and width of the $1s$ state of the kaonic hydrogen  and cross-section ratios mentioned in Eqs.~(\ref{1s_energy})~and~ (\ref{ratios}), as well as the poles found in the complex plane, we show in Fig.~\ref{fit2} the cross sections found with the parameter set labeled as Fit II. 
\begin{figure}[h!]
\includegraphics[width=0.368\textwidth]{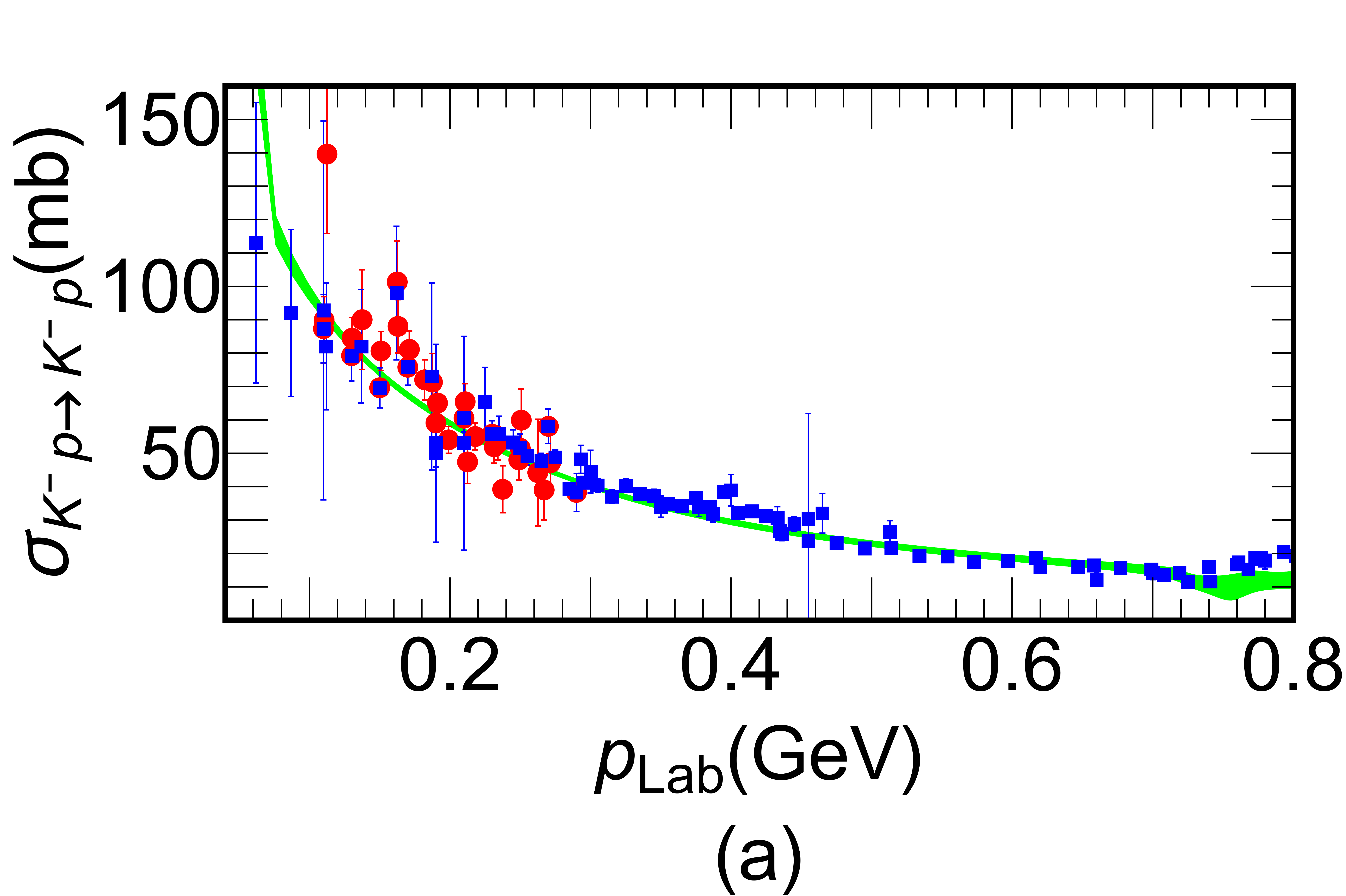}
\includegraphics[width=0.368\textwidth]{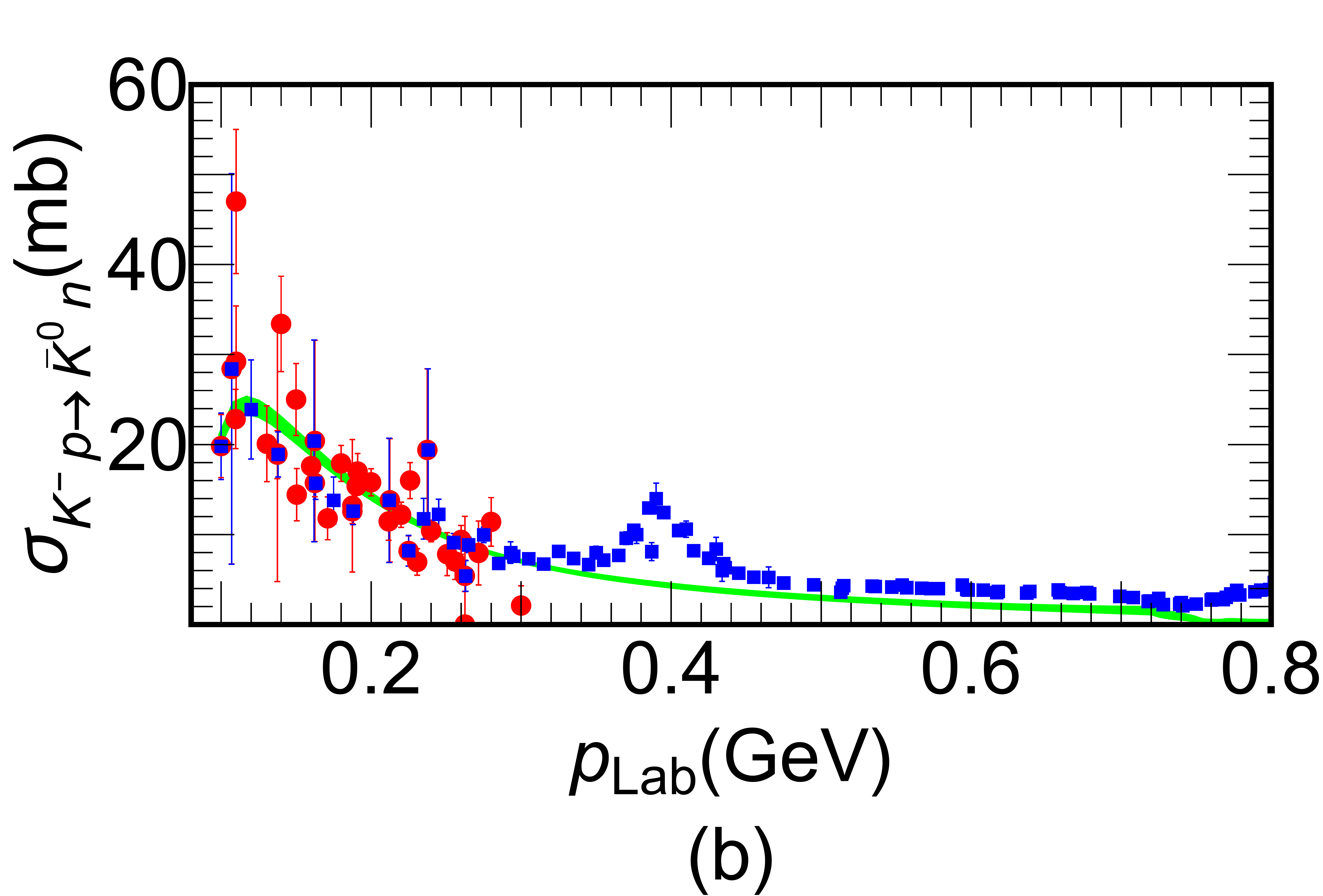}
\includegraphics[width=0.368\textwidth]{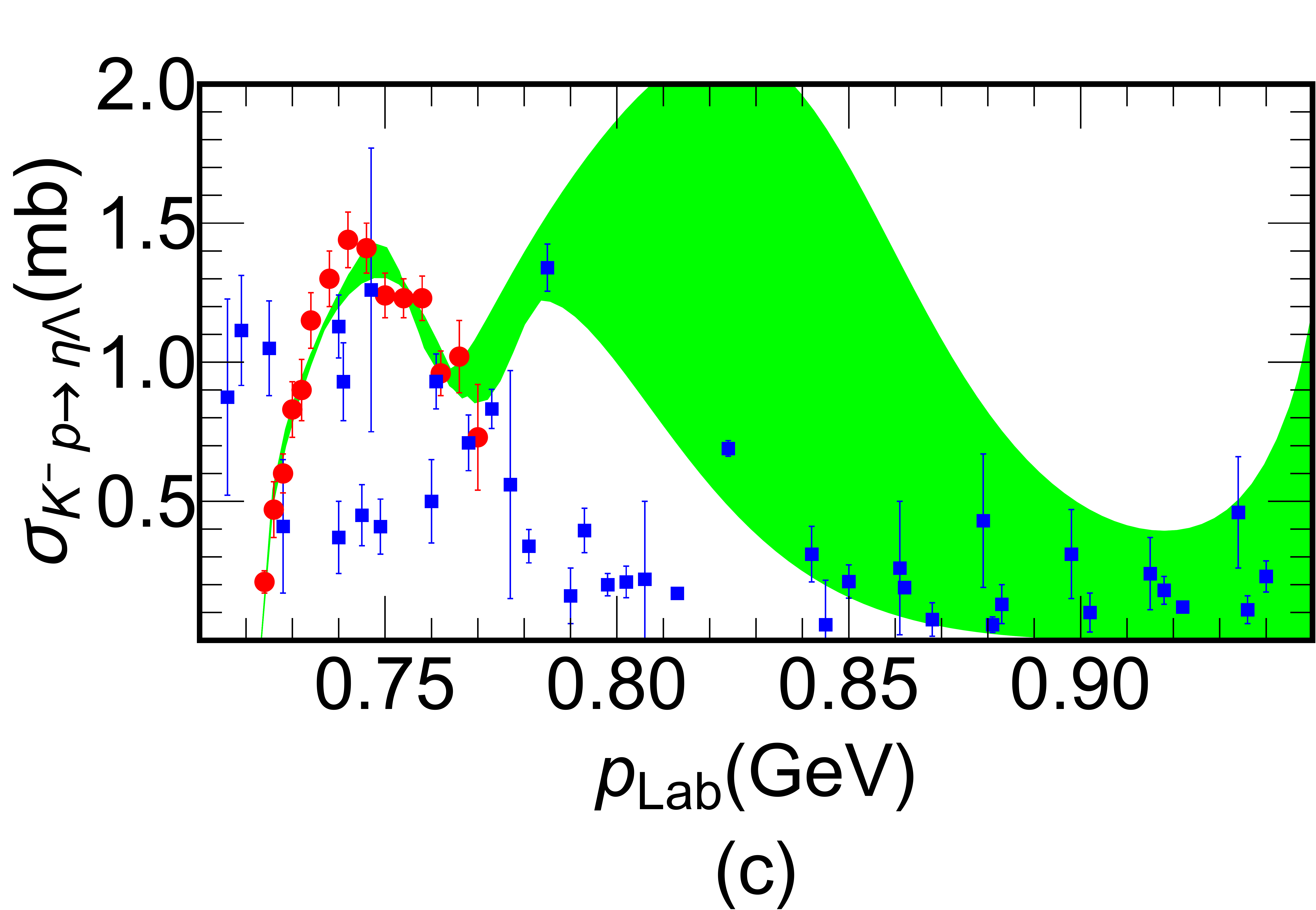}
\includegraphics[width=0.368\textwidth]{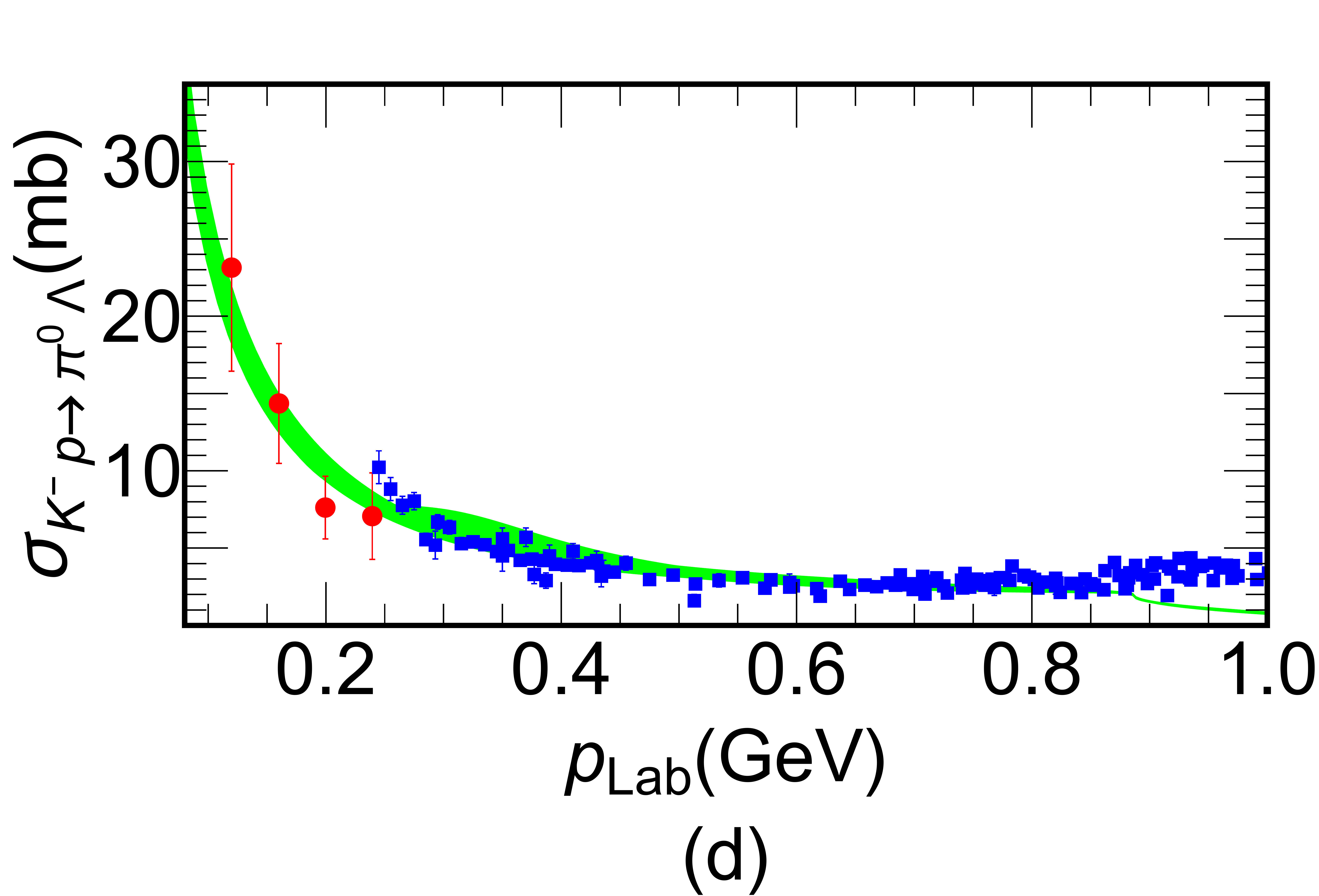}
\includegraphics[width=0.368\textwidth]{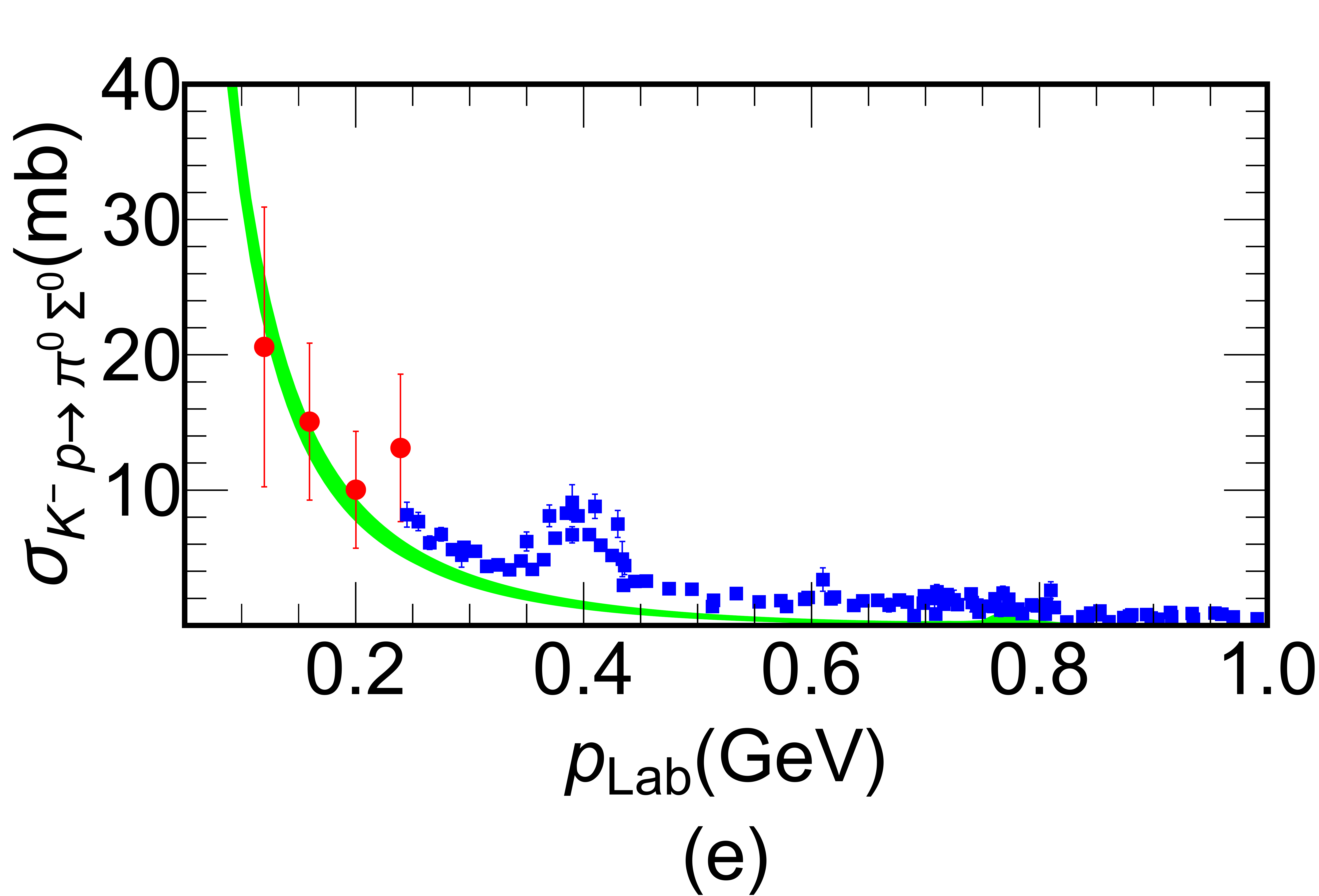}
\includegraphics[width=0.368\textwidth]{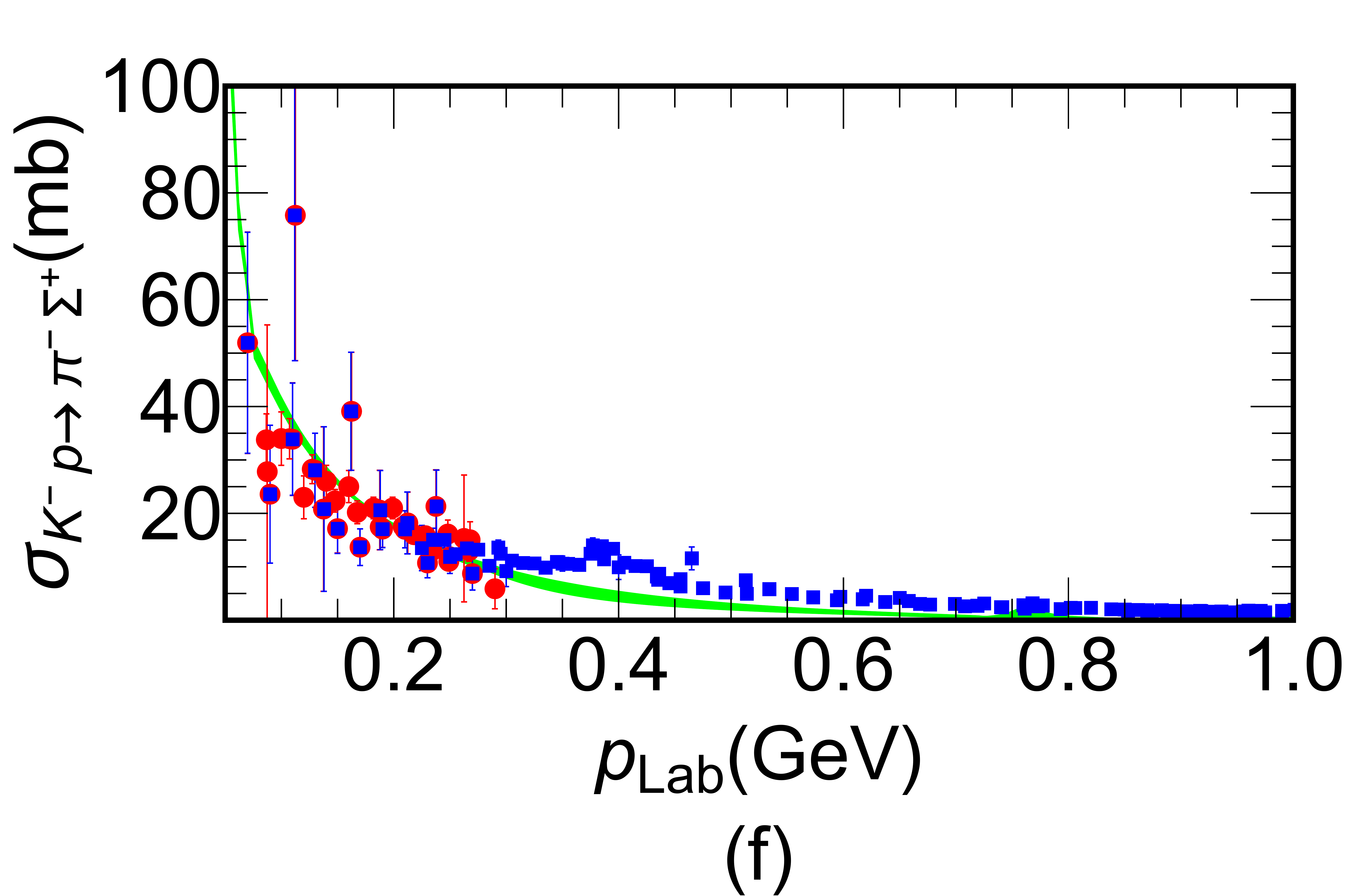}
\includegraphics[width=0.368\textwidth]{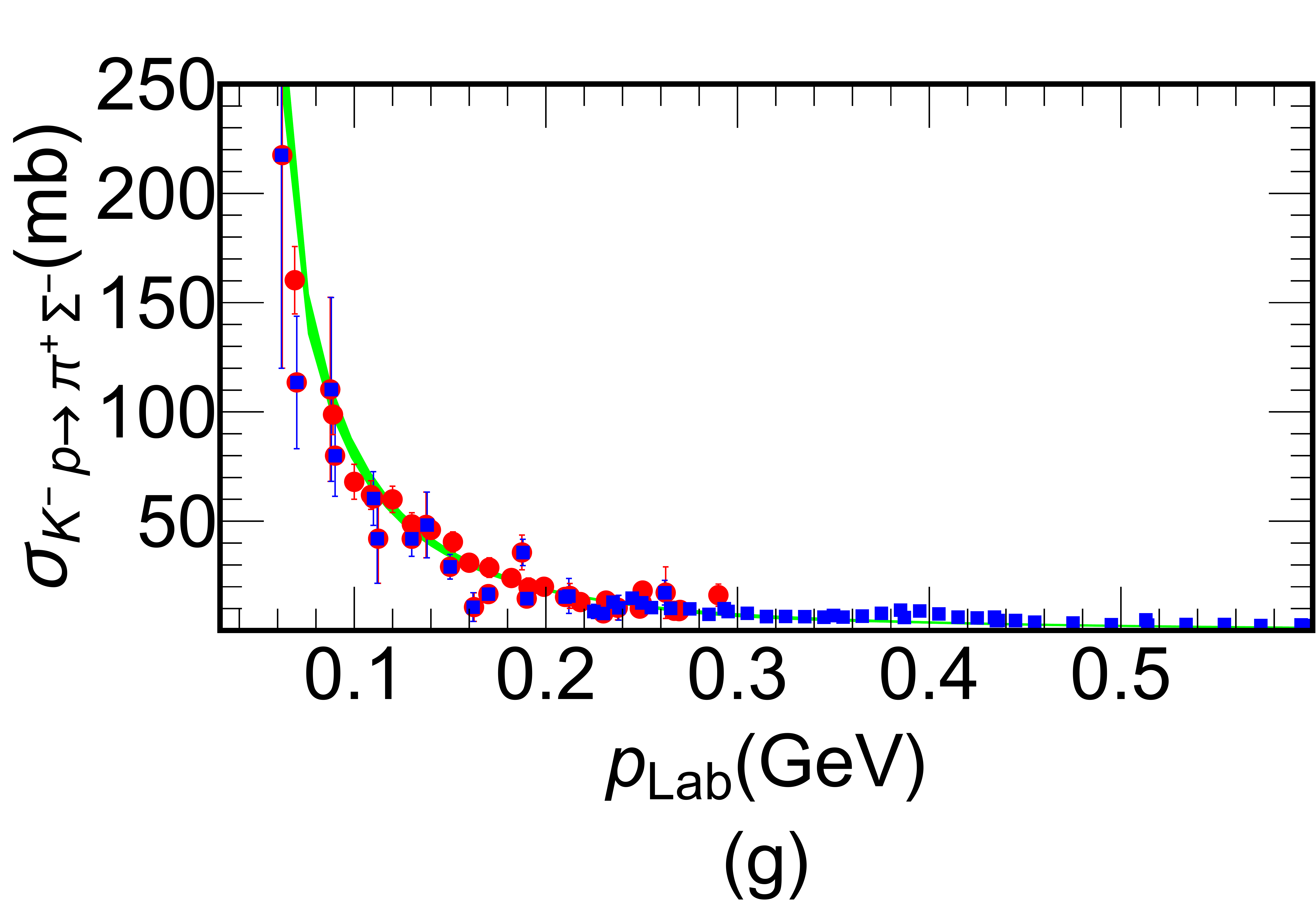}
\caption{Cross sections obtained with the parameter set Fit II given Table~\ref{par}. The data are taken from the same source as in Fig~\ref{fit1}.}\label{fit2}
\end{figure}
It can be seen that the cross sections are in good agreement with the data in the energy region corresponding to the filled circles (which are used to minimize the $\chi^2$), as expected, and the results stay near the data points at higher energies except for the case of $K^- p \to \eta \Lambda$. This finding may indicate, when comparing the two fits, that the results related to the poles found in the complex plane may be more reliable in the case of Fit~I, at energies beyond $\sim$~1.68 GeV (which corresponds to the laboratory momentum of about 0.77 GeV shown in Figs.~\ref{fit1}~and~\ref{fit2}). At lower energies, though, the two fits are of similar quality, implying that the poles obtained in amplitudes for both fits, in the complex plane, should be reliable at energy below $\sim 1.68$ GeV.  Besides this finding, the cross section ratios, as well as the energy shift of the $1s$ state of the kaonic hydrogen found within the two fits, as given in Table \ref{wr}, 
{\squeezetable\begin{table}[h!]
\caption{Results found for the energy shift and width of the $1s$ state of the kaonic hydrogen and the cross-section ratios defined in Eq.~(\ref{ratios}). The central value and errors  correspond to the mean value and the standard deviation, respectively, determined from the solutions satisfying Eq.~(\ref{criteria}).}\label{wr}
 \begin{tabular}{cccccc}
 \hline
&$\Delta E (\text{eV})$&$\Gamma(\text{eV})$&$\gamma$&$R_c$&$R_n$\\
\hline\hline
Fit I&$300\pm3$&$448\pm6$&$2.357\pm0.005$&$0.663\pm0.003$&$0.191\pm0.002$\\
Fit II&$301\pm 6$&$474\pm 17$&$2.364\pm0.008$&$0.668\pm0.003$&$0.193\pm0.002$\\
Data (from Refs.~\cite{Bazzi:2011zj,Tovee:1971ga,Nowak:1978au})&$283 \pm 36 \pm 6$&$549 \pm 89 \pm 22$&$2.36 \pm 0.12$&$0.664\pm0.033$&$0.189\pm0.015$\\\hline
\end{tabular}
\end{table}}
are in good agreement with the experimental data (see the values given in Sec.~\ref{sec:formalism}).  We, thus, find it useful to discuss the remaining results for both fits. 

Before continuing, though, a reader may wonder if having these fits at hand and having a good deal of overlap in the experimental data considered here and in Ref~\cite{Guo}, for instance, if it is possible to do a statistical comparison between the models used here and in Ref.~\cite{Guo}. A standard statistical test to compare models is the Fisher's test (F-test), though it can be used to compare nested models.  The formalism of the present work and the one in Ref.~\cite{Guo} cannot be treated as nested models. Another possible test, which can be used for nested as well as nonnested models is the Akaike-information-criterion (AIC).  Strictly speaking, though, an AIC  comparison is meaningful when exactly the same data is used in the fitting procedure and this condition is not satisfied here. For instance, the authors of Ref.~\cite{Guo}  consider the data on processes like $K^- p \to \pi^0 \pi^0 \Sigma^0$, $K^- p \to \pi^- \Sigma(1660)^+$ in the fit, which are not included in the present work. In such a situation, we could still adopt the following strategy: We could assume that the fits obtained in Ref.~\cite{Guo} for the processes considered in this work would not differ much from those the authors of Ref.~\cite{Guo} would have found by excluding the data not included here. In this way, we can calculate the AIC number as \cite{Sola:2016zeg} 
\begin{equation}
\text{AIC} = \chi^2_0 + \frac{2 n k}{k-n-1},
\end{equation}
with $n$ being the number of  parameters and $k$ the number of data points. Under such an assumption,  we find that the AIC number for the $O(p^2)$ fit of Ref.~\cite{Guo} is smaller than the AIC number for the $O(p)$ fit of Ref.~\cite{Guo} and the AIC number obtained for our model is smaller than the former two.  A lower value of AIC may be interpreted as a model more preferred by the data. However, due to the assumptions involved in comparing the models, any conclusions should be made with caution.

Going back to the results obtained, in Tables~\ref{poles01h}-\ref{poles13h}, we list the poles found in the complex plane, with the amplitudes obtained within both fits.  In the following subsections we also compare the properties of the states found in our analysis  with those available from other theoretical/experimental works. Before proceeding, though, we would like to discuss the procedure to calculate the $T$-matrix in the complex energy plane, which is needed to look for resonances/bound states formed in the systems under investigation.  For this, we calculate the loop function for the $i$th channel in the first (I) and second (II) Riemann sheet as \cite{Oller:1997ti,Jido:2003cb}: 
\begin{eqnarray}\nonumber
G_i(\sqrt{s}) = \left\{\begin{array}{cc}
G_i^{(I)}(\sqrt{s}), & {\rm for }~ {\rm Re}\{\sqrt{s}\}\!<\!(m_i\!+\!M_i)\\
&\\
G_i^{(II)}(\sqrt{s}), & {\rm for }~ {\rm Re}\{\sqrt{s}\} \ge \!(m_i\!+\!M_i)
\end{array},\right.
\end{eqnarray}
where
\begin{eqnarray}\label{ana_contG1}
G_i^{(I)}(\sqrt{s})&=& G_i(\sqrt{s}) \\\nonumber
G_i^{(II)}(\sqrt{s})&=& G_i^{(I)}(\sqrt{s}) -2\,i\,{\rm Im} \{G_i^{(I)}\} \\
&=& G_i^{(I)}(\sqrt{s}) + i \frac{M_i~ q_i^{(I)}}{2\pi \sqrt{s}}, \label{ana_contG2}
\end{eqnarray}
with $m_i$, $M_i$ being the masses of the $i$th-meson and baryon, and $q_i^{(I)}$ the center of mass momentum of the same channel on its first Riemann sheet (with a positive imaginary part).  If a pole appears in the complex plane, it can be seen in the complex amplitude for all the channels. Depending on the threshold of a given channel, the pole can appear below or above the threshold (i.e, on the corresponding first or second Riemann sheet of that channel).

\subsection{Isospin = 0, spin = 1/2}
In the case of  $I(J^P)=0~(1/2^-)$, in both types of fits, a double pole structure is found in the energy region around 1400 MeV (see Table~\ref{poles01h}), which can be related to $\Lambda(1405)$. The double pole nature of $\Lambda(1405)$ is widely discussed in the literature \cite{Roca:2017wfo,Kamiya:2016jqc,Oset:2015ksa,Borasoy:2005ie,GarciaRecio:2002td,Mai:2012dt,Jido:2003cb}. {\squeezetable\begin{table}[h!]
\caption{Pole positions and couplings of the $I(J^P)=0(1/2^-)$ states found. The central values and errors were obtained as explained in the caption of Table~\ref{par} (for the sake of space, the errors are represented as superscripts). Masses and widths are given in MeV.   The coupling of the state to a given channel are written as rows in the Table for Fit I and II ( the first (second) row is related to the results for Fit I (Fit II)). The symbol ``$-$'' indicates that we have ignored the states found with mass beyond 1680 MeV in Fit II}\label{poles01h}
\vspace{0.5cm}
\centering
\begin{tabular}{cccccc}
\hline\bigstrut[t]
&\multicolumn{2}{c}{$\Lambda(1405)$}&$\Lambda(1670)$&$\Lambda(1800)$\\
\hline\hline
Fit I&$\makecell{1373^{\pm5}-i\,114^{\pm9}}$&$\makecell{1426^{\pm1}-i\,16^{\pm1}}$&$\makecell{1681^{\pm1}-i\,16^{\pm2}}$&$\makecell{1734^{\pm7}-i\,19^{\pm2}}$&\\
\hline
Fit II&$\makecell{1385^{\pm5}-i\,124^{\pm10}}$&$\makecell{1426^{\pm1}-i\,15^{\pm2}}$&$\makecell{1681^{\pm2}-i\,7^{\pm1}}$&$\makecell{-}$\\
\hline\hline
\multirow{2}{*}{$\bar KN$}&$0.84^{\pm0.14}-i\,1.91^{\pm0.06}$& $2.44^{\pm0.05}+i\,0.69^{\pm0.08}$& $0.33^{\pm0.02}-i\,0.38^{\pm0.03}$&$0.14^{\pm0.05}-i\,0.12^{\pm0.07}$ \\
& $0.66^{\pm0.35}-i\,1.93^{\pm0.12}$& $2.43^{\pm0.16}+i\,0.63^{\pm0.23}$& $0.15^{\pm0.06}-i\,0.19^{\pm0.13}$&$-$  \\
\multirow{2}{*}{$K\Xi$}& $-0.51^{\pm0.05}+i\,0.49^{\pm0.06}$& $0.59^{\pm0.09}-i\,0.19^{\pm0.04}$&$2.74^{\pm0.26}+i\,0.25^{\pm0.22}$ &$1.26^{\pm0.60}-i\,0.39^{\pm0.28}$  \\
& $-0.55^{\pm0.13}+i\,0.27^{\pm0.06}$& $0.72^{\pm0.14}-i\,0.14^{\pm0.08}$&$0.33^{\pm0.64}+i\,0.28^{\pm0.34}$ &$-$  \\
\multirow{2}{*}{$\pi \Sigma$}&  $-2.04^{\pm0.07}+i\,2.29^{\pm0.08}$&$-0.87^{\pm0.06}-i\,1.05^{\pm0.09}$ & $0.27^{\pm0.02}+i\,0.42^{0.06}$&$0.09^{\pm0.05}-i\,0.14^{\pm0.07}$ \\
 & $-2.05^{\pm0.11}+i\,2.27^{\pm0.09}$&$-0.90^{\pm0.08}-i\,0.96^{\pm0.15}$ & $-0.11^{\pm0.20}-i\,0.13^{\pm0.35}$&$-$ \\
\multirow{2}{*}{$\eta\Lambda$}& $-0.71^{\pm0.07}-i\,1.24^{\pm0.04}$&$2.45^{\pm0.05}+i\,0.21^{\pm0.04}$ &$-0.83^{\pm0.14}+i\,0.11^{\pm0.08}$ & $-0.50^{\pm0.23}+i\,0.49^{\pm0.24}$\\
& $-0.80^{\pm0.10}-i\,1.24^{\pm0.06}$&$2.34^{\pm0.13}+i\,0.16^{\pm0.04}$ &$-0.19^{\pm0.10}-i\,0.20^{\pm0.06}$ & $-$ \\
\multirow{2}{*}{$\bar K^*N$}& $0.86^{\pm0.08}-i\,0.04^{\pm0.10}$&$-0.16^{\pm0.10}+i\,0.26^{\pm0.03}$ &$-0.18^{\pm0.08}-i\,0.05^{\pm0.03}$ &$-0.15^{\pm0.11}+i\,0.05^{\pm0.04}$ \\
& $0.62^{\pm0.28}-i\,0.18^{\pm0.14}$&$0.04^{\pm0.36}+i\,0.23^{\pm0.19}$ &$0.50^{\pm0.92}+i\,0.01^{\pm0.10}$ &$-$  \\
\multirow{2}{*}{$K^*\Xi$}& $1.23^{\pm0.11}-i\,0.08^{\pm0.09}$&$-0.36^{\pm0.12}+i\,0.42^{\pm0.05}$ & $-2.05^{\pm0.25}+i\,0.22^{\pm0.13}$&$1.01^{\pm0.47}+i\,0.22^{\pm0.18}$ \\
& $1.17^{\pm0.12}-i\,0.40^{\pm0.12}$&$0.00^{\pm0.19}+i\,0.44^{\pm0.08}$ & $1.04^{\pm2.99}-i\,0.19^{\pm0.30}$&$-$  \\
\multirow{2}{*}{$\rho\Sigma$}& $0.16^{\pm0.11}+i\,0.29^{\pm0.07}$&$-0.24^{\pm0.09}-i\,0.01^{\pm0.02}$ &$0.23^{\pm0.16}-i\,0.09^{\pm0.08}$ &$-0.28^{\pm0.28}-i\,0.04^{\pm0.03}$ \\
& $0.57^{\pm0.24}+i\,0.41^{\pm 0.19}$&$-0.47^{\pm0.43}+i\,0.03^{\pm0.18}$ &$-1.76^{\pm2.58}+i\,0.10^{\pm0.37}$ &$-$ \\
\multirow{2}{*}{$\omega\Lambda$}& $-0.26^{\pm0.03}+i\,0.28^{\pm0.03}$&$-0.37^{\pm0.02}-i\,0.15^{\pm0.02}$ & $0.51^{\pm0.06}-i\,0.09^{\pm0.03}$&$-0.32^{\pm0.15}-i\,0.07^{\pm0.06}$ \\
& $-0.23^{\pm0.10}+i\,0.33^{\pm0.06}$&$-0.45^{\pm0.09}-i\,0.16^{\pm0.07}$ & $-0.32^{\pm0.71}+i\,0.05^{\pm0.08}$&$-$  \\
\multirow{2}{*}{$\phi\Lambda$}& $0.46^{\pm0.07}-i\,0.44^{\pm0.06}$&$0.62^{\pm0.05}+i\,0.25^{\pm0.03}$ &$-0.66^{\pm0.10}+i\,0.12^{\pm0.04}$ &$0.39^{\pm0.19}+i\,0.11^{\pm0.07}$ \\
& $0.44^{\pm0.27}-i\,0.58^{\pm0.13}$&$0.82^{\pm0.30}+i\,0.29^{\pm0.19}$ &$0.60^{\pm1.19}-i\,0.09^{\pm0.14}$ &$-$ \\\hline
\end{tabular}
\end{table}}
Our results are compatible with the pole values obtained in these former works, as well as with those determined by the CLAS collaboration \cite{Lu:2013nza} from the data on the electroproduction  of $\Lambda(1405)$, with the lower mass pole being near 1368 MeV and the higher mass pole near 1423 MeV. The pole values obtained from fits constrained by photoproduction data are also worth mentioning, for instance, those obtained in Refs.~\cite{Roca:2013cca,Mai:2014xna}. The best solution in the former work corresponds to the poles for $\Lambda(1405)$: $1429^{+8}_{-7} - i 12^{+2}_{-3}$ and $1325^{+15}_{-15} - i 90^{+12}_{-18}$ MeV.  In the latter one  (\cite{Roca:2013cca}), a fit to photoproduction data is made and the best solution is found to reasonably describe the data on cross sections for different final states in $K^- p$ collisions. The corresponding poles obtained in  Ref.~\cite{Roca:2013cca} are: $1352 - i 48$ MeV and $1419 - i 29$ MeV. Our findings  agree better with those in Ref.~\cite{Mai:2014xna}.

We give the couplings of these poles to the different meson-baryon channels considered in the present work in Table~\ref{poles01h}. The coupled channel treatment of pseudoscalar-baryon and vector-baryon systems is a particular feature of our formalism and it allows us to obtain the information on the coupling of the low lying resonances, like, $\Lambda(1405)$, to both type of channels. The information on the coupling of the states to vector-baryon channels is useful in studies of processes like the photoproduction/electroproduction of $\Lambda(1405)$ through a $t$-channel  vector exchange (as done in Refs.~\cite{Nam:2008jy,Kohri:2009xe,Kim:2012pz,Nam:2006cx,Lu:2013nza,Nakamura:2013boa}}).

Table~\ref{poles01h} also shows a pole around 1680 MeV, which is related to $\Lambda(1670)$. The mass and width of this state range, according to the particle data group (PDG)~\cite{pdg}, between 1670$-$1680 MeV and 25$-$50 MeV, respectively. The pole position found with Fit I: $(1681 \pm1) - i (16\pm 2)$ MeV is in better agreement with the properties of $\Lambda(1670)$ from the PDG~\cite{pdg}.  We have determined the branching ratios of this state for channels $\bar K N$, $\pi \Sigma$, and $\eta \Lambda$ and find them, respectively, to be  28$\%$, 34$\%$, and 25$\%$ with the central values of the parameters in Fit I and  19$\%$, 61$\%$, and 7$\%$ with the central values in Fit II (given in Table~\ref{par}). The former values are in better agreement  with the values: 20-30$\%$, 25-55$\%$, and  10-25$\%$ given in Ref.~\cite{pdg}. This finding is in line with the earlier discussions on the reliability of the results obtained within Fit II beyond $\sim$1680 MeV,  due to the disagreement of the $K^- p \to \eta \Lambda$ cross sections at energies $\gtrsim$ 1680 MeV. With this finding at hand, and with fits shown in Figs.~\ref{fit1} and  \ref{fit2}, we do not discuss the properties of the states with mass $\geq$ 1680 MeV found within Fit II.  As mentioned earlier, however, the two fits are of similar quality for energies $\leq$ 1680 MeV and we, thus,  continue discussing the properties of states found with both the fits when the mass is lower than  $\sim$ 1680 MeV.

 In view of the results found in our work, and as widely accepted, both $\Lambda(1405)$ and $\Lambda(1670)$ can be interpreted as states arising from pseudoscalar-baryon dynamics. We find a pole with $I(J^P)=0~(1/2^-)$, which gets contribution from vector-baryon dynamics as well, with mass around 1730 MeV in Fit I.  
 Only one $1/2^-$ $\Lambda$ state is listed in this energy region  by the PDG~\cite{pdg}, after $\Lambda(1670)$ ,  which encompasses evidences on $I(J^P)=0~(1/2^-)$ states with masses ranging from 1720$-$1850 MeV and widths ranging over 100$-$600 MeV. It is then quite possible that more than one state get classified under the same label $\Lambda(1800)$. From our study and in  light of the information available from the PDG~\cite{pdg}, it can be said that a state is found around 1730 MeV with a width around 40 MeV.  However, missing channels not considered in the present work could have an impact on the width of this state and make it larger. A more detailed study involving such channels and considering data on reactions producing  $VB$ channels should be done in future to investigate further properties of this state.

A comment regarding the widths of the states found in our work is here in order. The half widths of the states with mass around or above $1800$ MeV have been determined from the real axis (by reading the full width at the half maximum of the related peaks appearing in the squared amplitudes, on the real axis, a quite common procedure in this kind of problem~\cite{Oset:2012ap}). This is done because the widths of the vector mesons, here and throughout the work, are not taken into account when calculating the amplitudes in the complex plane, since such a consideration  would imply a variable mass of the vector meson and, hence, a not well defined branch cut in the complex plane. However, as explained earlier, the amplitudes on the real axis have been obtained by taking the finite widths of the vector mesons into account. Thus, a better estimation of the widths of the resonances is obtained from the real axis. In such cases, the couplings of the states to the different channels are also determined from the real axis.

\subsection{Isospin = 1, spin = 1/2}
In the case of  $1/2^-$ isovector scattering amplitudes studied in the complex plane,  two poles appear around 1400 MeV with the parameter set Fit I (see Table~\ref{poles11h}),
{\squeezetable
\begin{table}[h!]
\caption{Pole positions and couplings of the $I(J^P)=1(1/2^-)$ states found in our work. The central values and errors were obtained as explained in the caption of Table~\ref{par} (for the sake of space, the errors are represented as superscripts).}\label{poles11h}
\vspace{0.5cm}
\centering
\begin{tabular}{ccccc}
\hline\bigstrut[t]
&\multicolumn{2}{c}{$\Sigma$'s around 1400 MeV}&$\Sigma(1620)$ or $\Sigma(1670)$ &$\Sigma(1900)$\\
\hline\hline
Fit I&$\makecell{1396^{\pm1}-i\,5^{\pm2}}$&$\makecell{1367^{\pm24}-i\,57^{\pm21}}$&$\makecell{1630^{\pm33}-i\,104^{\pm13}}$&$\makecell{1853^{\pm10}-i\,150^{\pm10}}$\\
Fit II&$-$&$\makecell{1399^{\pm35}-i\,36^{\pm9}}$&$-$&$-$\\
\hline\hline
\multirow{2}{*}{$\bar KN$}&$0.18^{\pm0.03}+i\,0.14^{\pm0.05}$&$0.08^{\pm0.48}+i\,0.52^{\pm 0.73}$&$1.47^{\pm0.08}-i\,0.017^{\pm0.07}$&$-0.86^{\pm0.03}+i\,0.79^{\pm0.02}$\\
&$-$& $0.50^{\pm0.29}+i\,0.33^{\pm0.18}$& $-$& $-$\\
\multirow{2}{*}{$K\Xi$}&$1.06^{\pm0.22}+i\,1.45^{\pm0.12}$&$0.62^{\pm0.47}-i\,0.42^{\pm1.00}$&$2.89^{\pm0.26}-i\,0.65^{\pm0.24}$&$0.84^{\pm0.03}-i\,0.39^{\pm0.05}$\\
&$-$& $0.81^{\pm0.42}+i\,0.41^{\pm0.15}$& $-$&$-$  \\
\multirow{2}{*}{$\pi \Sigma$}&$-0.17^{\pm0.09}-i\,020^{\pm0.03}$&$0.77^{\pm0.96}-i\,0.67^{\pm1.22}$&$0.71^{\pm0.33}-i\,1.63^{\pm0.19}$&$-0.02^{\pm0.04}+i\,0.32^{\pm0.08}$\\
&$-$& $1.08^{\pm0.12}+i\,0.19^{\pm0.21}$&$-$ & $-$\\
\multirow{2}{*}{$\pi\Lambda$}&$0.03^{\pm0.10}+i\,0.07^{\pm0.06}$&$-0.91^{\pm1.32}+i\,0.39^{\pm0.81}$&$-0.26^{\pm0.34}-i\,0.23^{\pm0.18}$&$0.36^{\pm0.2}+i\,1.54^{\pm0.04}$\\
&$-$& $-1.40^{\pm0.18}-i\,0.07^{\pm0.10}$&$-$ &$-$  \\
\multirow{2}{*}{$\eta\Sigma$}&$-0.43^{\pm0.03}-i\,0.23^{\pm0.09}$&$0.31^{\pm0.31}-i\,0.59^{\pm1.12}$&$-2.14^{\pm0.24}-i\,0.13^{\pm0.11}$&$0.07^{\pm0.03}-i\,0.43^{\pm0.02}$\\
&$-$& $0.27^{\pm0.10}-i\,0.19^{\pm0.11}$&$-$ &$-$ \\
\multirow{2}{*}{$\bar K^* N$}&$0.04^{\pm0.10}+i\,0.15^{\pm0.07}$&$-1.69^{\pm1.99}+i\,0.31^{\pm0.68}$&$-0.31^{\pm0.09}-i\,0.11^{\pm0.16}$&$0.71^{\pm0.05}-i\,0.05^{\pm0.02}$\\
&$-$ & $-3.46^{\pm0.21}-i\,0.06^{\pm0.15}$&$-$ & $-$\\
\multirow{2}{*}{$K^*\Xi$}&$-0.50^{\pm0.22}-i\,0.38^{\pm0.08}$&$1.40^{\pm2.11}-i\,1.10^{\pm2.38}$&$-1.80^{\pm0.47}-i\,0.37^{\pm0.14}$&$-0.98^{\pm0.14}-i\,0.72^{\pm0.06}$\\
&$-$& $-0.01^{\pm0.59}-i\,0.21^{\pm0.08}$&$-$ &$-$ \\
\multirow{2}{*}{$\rho\Sigma$}&$-0.15^{\pm0.07}-i\,0.14^{\pm 0.04}$&$0.76^{\pm1.02}-i\,0.58^{\pm0.85}$&$-0.76^{\pm0.18}-i\,0.53^{\pm0.49}$&$-1.10^{\pm0.04}-i\,0.34^{\pm0.03}$\\
&$-$& $3.60^{\pm0.61}-i\,0.69^{\pm0.16}$&$-$ & $-$\\
\multirow{2}{*}{$\rho\Lambda$}&$0.36^{\pm0.18}+i\,0.29^{\pm0.07}$&$-0.95^{\pm1.50}+i\,0.93^{\pm1.84}$&$2.44^{\pm0.50}+i\,0.94^{\pm0.27}$&$1.51^{\pm0.25}+i\,0.82^{\pm0.09}$\\
&$-$& $-1.26^{\pm0.19}+i\,0.09^{\pm0.07}$&$-$ &$-$ \\
\multirow{2}{*}{$\omega\Sigma$}&$-0.15^{\pm0.11}-i\,0.14^{\pm0.05}$&$1.03^{\pm1.35}-i\,0.55^{\pm1.10}$&$-0.14^{\pm0.23}-i\,0.44^{\pm0.14}$&$-0.64^{\pm0.10}-i\,0.23^{\pm0.04}$\\
&$-$& $2.15^{\pm0.20}-i\,0.13^{\pm0.09}$&$-$ &$-$ \\
\multirow{2}{*}{$\phi\Sigma$}&$0.27^{\pm0.17}+i\,0.24^{\pm0.08}$&$-1.73^{\pm2.27}+i\,0.90^{\pm1.82}$&$0.42^{\pm0.38}+i\,0.53^{\pm0.24}$&$1.04^{\pm0.20}+i\,0.39^{\pm0.07}$\\
&$-$ & $-3.23^{\pm0.39}+i\,0.20^{\pm0.11}$&$-$ &$-$\\\hline
\end{tabular}
\end{table}}
while only one pole is obtained with Fit II. It can be seen from Figs.~\ref{fit1}~and~\ref{fit2} and the results in Table~\ref{wr} that  the quality of both fits is similar in the energy region near the threshold. Thus, from our work, it is difficult to distinguish the possibility of the existence of one or two isovector poles around 1400 MeV. But even if two poles exist in nature, they may be related to the same state due to the proximity of the masses and the widths.  Thus, it can be concluded that a $\Sigma$ state does seem to appear in this energy region. It should be useful to compare our findings with those available in the literature. As mentioned in the introduction of this article, the information on the light $\Sigma$'s is less abundant when compared to light $\Lambda$'s. Still, we can compare our results with other works \cite{Oller:2000fj,Guo,Moriya:2013eb,Wu:2009tu,Wu:2009nw,Gao:2010hy,Xie:2014zga,Xie:2017xwx}, where almost all agree on the existence of one $\Sigma$ state around 1380 MeV with the width of about 60 MeV. An evidence for two poles around 1400 MeV, in isospin 1 amplitudes, has been discussed in Refs.~\cite{Oller:2000fj,Guo,Moriya:2013eb}, out of which Ref.~\cite{Guo} finds one of the poles to be narrower, as is the case of the results for Fit I listed in Table~\ref{poles11h}. Suggestions have been made to find this state in the $\chi_{c0}$ decay into $\bar \Sigma \Sigma\pi$~\cite{Wang:2015qta} and in the $\Lambda^+_c$ decay into $\eta\pi^+\Lambda$~\cite{Xie:2017xwx}. We would like to mention a couple of facts related to these lowlying $\Sigma$ states: (1) their masses lie in the energy region where the left hand cut is crossed for some coupled channels. (2) Though more rigorous treatments of the $u$-channel amplitude should be considered in future to obtain the nonperturbative $T$ matrix, we find, interestingly, that  the second pole in Fit I as well as the pole in Fit II are found to continue to appear in the complex plane even if the contribution from the $u$-channel diagrams is switched off. 

In the energy region where vector-baryon thresholds are open,  two poles with Fit I are found in the energy region 1600$-$1900 MeV, which  can be related  to $\Sigma(1620)$ and $\Sigma(1900)$, respectively, listed by the PDG~\cite{pdg}. Actually, we have studied the possibility of relating the state at $1630 \pm 33 -i(104\pm13)$ MeV to $\Sigma(1670)$ as well as $\Sigma(1620)$. Little is known about both these $\Sigma$s and the PDG \cite{pdg} indicates that each of them may be related to two states, of which the spin-parity of only one (in each case) is known. The spin-parity  of one of the $\Sigma(1620)$s is given as $1/2^-$ by the PDG and for one of the $\Sigma(1670)$s as $3/2^-$~\cite{pdg}.  For a better analysis, we study the following decay ratios known for $\Sigma(1670)$ (with unknown spin-parity)~\cite{pdg},
\begin{eqnarray} 
& ~~~~~~~\dfrac{\Gamma (\Sigma(1670) \to \bar K N)}{\Gamma (\Sigma(1670) \to \pi \Sigma)} < 0.75,  & \\
&0.05  \lesssim \dfrac{\Gamma (\Sigma(1670) \to \pi\Lambda)}{\Gamma (\Sigma(1670) \to \pi \Sigma)} \lesssim 0.85,&
\end{eqnarray}
and from the state in our Fit I, the former ratio is obtained to be $\sim$0.5 and the latter one is found $\sim$0.06. In the case of $\Sigma(1620)~(1/2^-)$, the following partial widths are known from different partial-wave analyses~\cite{pdg}:
\begin{eqnarray}
&0.08 < \dfrac{\left(\Gamma(\Sigma(1620) \to \bar K N) \Gamma(\Sigma(1620) \to \pi \Sigma)\right)^{1/2}}{\Gamma_{\text{total}} }< 0.35, &\\
&0.1 < \dfrac{\left(\Gamma(\Sigma(1620) \to {\bar K N}) \Gamma(\Sigma(1620) \to \pi \Lambda)\right)^{1/2}}{\Gamma_{\text{total}} }< 0.15,&  \\
&0.08 < \dfrac{\Gamma(\Sigma(1620)\to \bar K N)} {\Gamma_{\text{total}} }< 0.35, &  
\end{eqnarray}
and we obtain them to be 0.37,  0.10, and 0.26, respectively. This analysis shows that our state can be associated to $\Sigma(1620) (1/2^-)$ as well as to $\Sigma(1670)$ with unknown spin-parity, which, in turn, may imply that both these states are not different. It may be useful to give the branching ratios of our state $1630 \pm 33 -i(104\pm13)$ MeV here. We find that decay ratios to $\bar KN$, $\pi \Sigma$, $\pi \Lambda$, $\eta \Sigma$ and $K \Xi$ are 26.3$\%$, 52.2$\%$, 3.5$\%$, 7.9$\%$ and 7.6$\%$, respectively. Not much is known about $ \Sigma(1900)$ either, it has been found in the partial-wave analyses of Refs.~\cite{zhang,Kamano:2015hxa}. The mass and width in Ref.~\cite{pdg} of $\Sigma(1900)$ are in agreement with those in Table~\ref{poles11h}. 

\subsection{Isospin = 0, spin = 3/2}
The vector-baryon systems can have a total spin 1/2 or 3/2 in $s$-wave interactions. Thus, we can study states with  spin-parity $(J^P) = (3/2^-)$ too. Such states arise purely from vector-baryon dynamics.  In the case of the $I (J^P)=0(3/2^-)$ configuration, we find a state with mass around 1800 MeV,  in fit I (see Table~\ref{poles03h}). 
\begin{table}[h!]
\caption{Pole positions and couplings of the $I(J^P) = 0~(3/2^-)$ states found. The central values and errors were obtained as explained in the caption of Table~\ref{par}. Since $PB$ systems in $s$-wave can
only have $J^P=1/2^-$, there is no coupling between the states listed in this table and the $PB$ channels in our model. Masses and widths are in MeV. The width gets contribution from the widths of the vector mesons (see text for more details). }\label{poles03h}
\vspace{0.5cm}
\centering
\begin{tabular}{ccc}
\hline
\bigstrut[t]
&$\Lambda(1690)$\\
~~~~~Fit I~~~~~&$1802^{\pm 7}-i\, 1.3^{\pm 0.8}$\\
\hline\hline
$\bar K^*N$& $0.91^{\pm0.03}+i\,0.017^{\pm0.001}$ \\
$K^*\Xi$& $3.30^{\pm0.05}+i\,0.061^{\pm0.002}$  \\
$\rho\Sigma$& $0.51^{\pm0.35}+i\,0.010^{\pm0.007}$\\
$\omega\Lambda$&$-0.06^{\pm0.03}-i\,0.001^{\pm0.001}$\\
$\phi\Lambda$& $0.60^{\pm0.06}+i\,0.011^{\pm0.001}$ \\\hline
\end{tabular}
\end{table}
To associate this state with a known $\Lambda$,  we look for known $3/2^-$ states listed by the PDG~\cite{pdg}, and find that there are two such $\Lambda$s in 1690-2050~MeV: $\Lambda(1690)$, with mass and width of $1697\pm6$ MeV and $65\pm 14$ MeV, respectively, and $\Lambda(2050)$,  with mass and width listed as $2056\pm 22$ and $493\pm 60$ MeV, respectively, out of which the latter one has been catalogued in Ref.~\cite{pdg}, so far,  motivated only by the partial-wave analysis of  $\bar K N$ multichannel reactions done in Ref.~\cite{zhang}. A full comparison is difficult in this case, since in our formalism, the $J^P=3/2^-$ VB channels do not couple to $J^P=1/2^-$ $PB$ channels in $s$-wave. The small widths of the states given in Table~\ref{poles03h} are due to the finite widths of the vector mesons involved in the dynamics. For a more reliable determination of the widths, PB and VB channels should be coupled in this sector too, including other mechanisms, like those in Ref.~\cite{Garzon:2012np} and including decuplet baryons in our formalism.  In addition to this, reactions involving VB final states might be included in the set of data fitted in the analysis. Such extensions of our work should be done in future. 

\subsection{Isospin = 1, spin = 3/2}
Some states, with $I(J^P)=1~(3/2^-)$, are also  found in our work, as shown in Table~\ref{poles13h}. 
\begin{table}[h!]
\caption{Pole positions and couplings of the $I(J^P)=1(3/2^-)$ states found in our work. The central values and errors were obtained as explained in the caption of Table~\ref{par}. Since $PB$ systems in s-wave can only have $J^P=1/2^-$, there is no coupling between the states listed in this table and the $PB$ channels (in our model).}\label{poles13h}
\vspace{0.5cm}
\centering
\begin{tabular}{cc}
\hline\bigstrut[t]
&$\Sigma(1670)$\\
\hline\hline
~~~~~~~~Fit I&$\makecell{1617^{\pm37}-i\,2^{\pm 1}}$\\
\hline\hline
$\bar K^*N$~& $0.41^{\pm0.13}+i\,0.003^{\pm0.015}$  \\
$K^*\Xi$~& $3.84^{\pm1.48}+i\,0.14^{\pm0.19}$  \\
$\rho\Sigma$~& $0.44^{\pm0.22}+i\,0.03^{\pm0.07}$\\
$\rho\Lambda$~& $-1.02^{\pm0.43}-i\,0.04^{\pm0.06}$ \\
$\omega\Sigma$&$-1.25^{\pm0.52}-i\,0.05^{\pm0.07}$  \\
$\Phi\Sigma$~&$2.59^{\pm1.01}+i\,0.10^{\pm0.13}$\\\hline
\end{tabular}
\end{table}
With Fit I, a pole is found around 1617 MeV which can be associated with the $3/2^-$ $\Sigma(1640)$~\cite{pdg}, whose mass and width range in the interval $1669\pm 7$ MeV and $64^{+10}_{-14}$, respectively. 
As mentioned earlier, in our model there is no coupling between the PB and VB channels in the spin 3/2 configuration, and, thus, the states get small widths owing to the instability of the vector mesons, which is taken into account by calculating the loop functions as in Eq.~(\ref{Gconv}).  For a better estimation of the widths, it may be important to consider transitions from vector-baryon to pseudoscalar-baryon channels in spin 3/2 too, but it is beyond the scope of the present work. 


\subsection{Additional information for  the $K^-p\to K^-p$ reaction}
\begin{figure}[h!]
\centering
\includegraphics[width=0.48\textwidth]{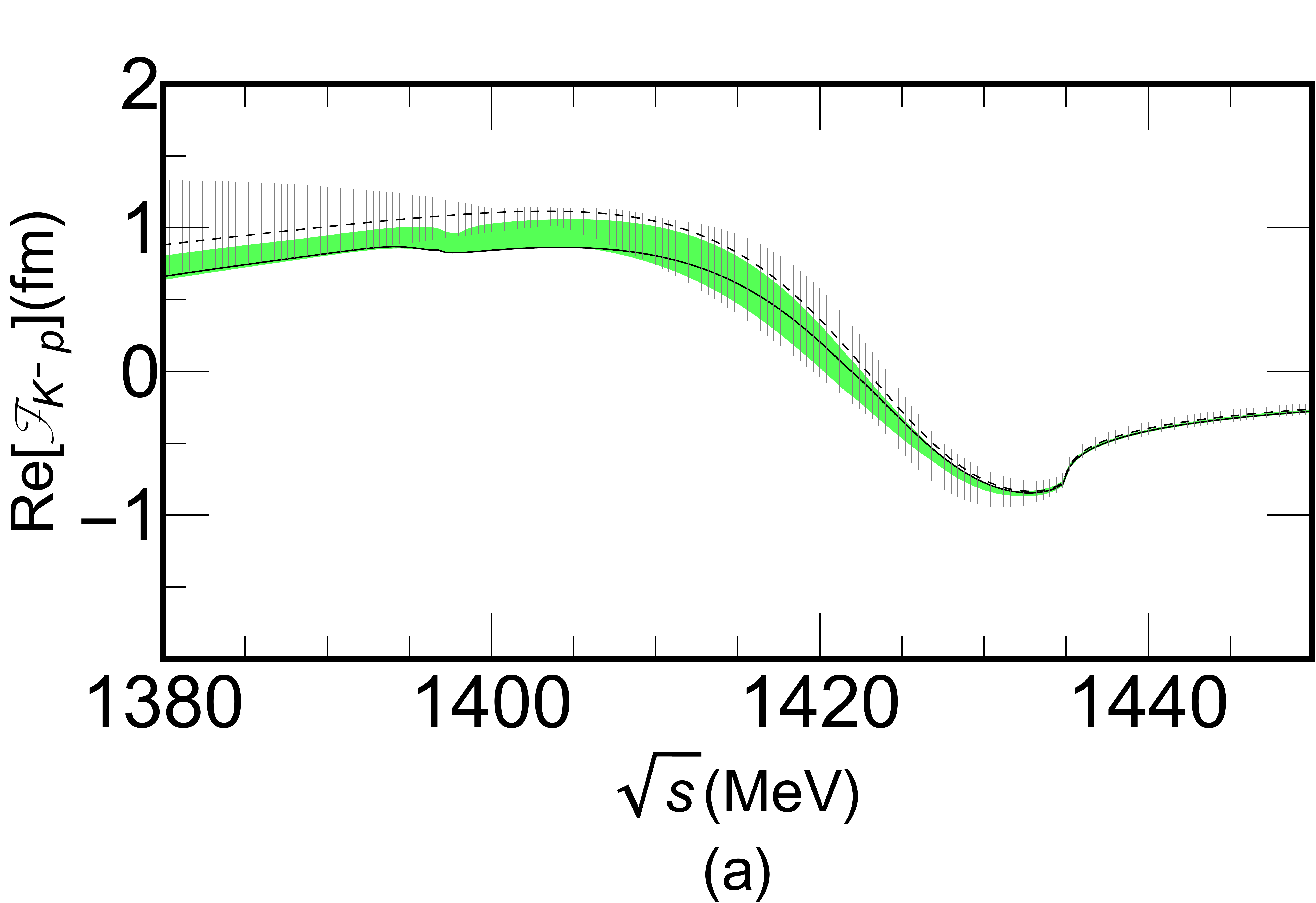}\quad
\includegraphics[width=0.48\textwidth]{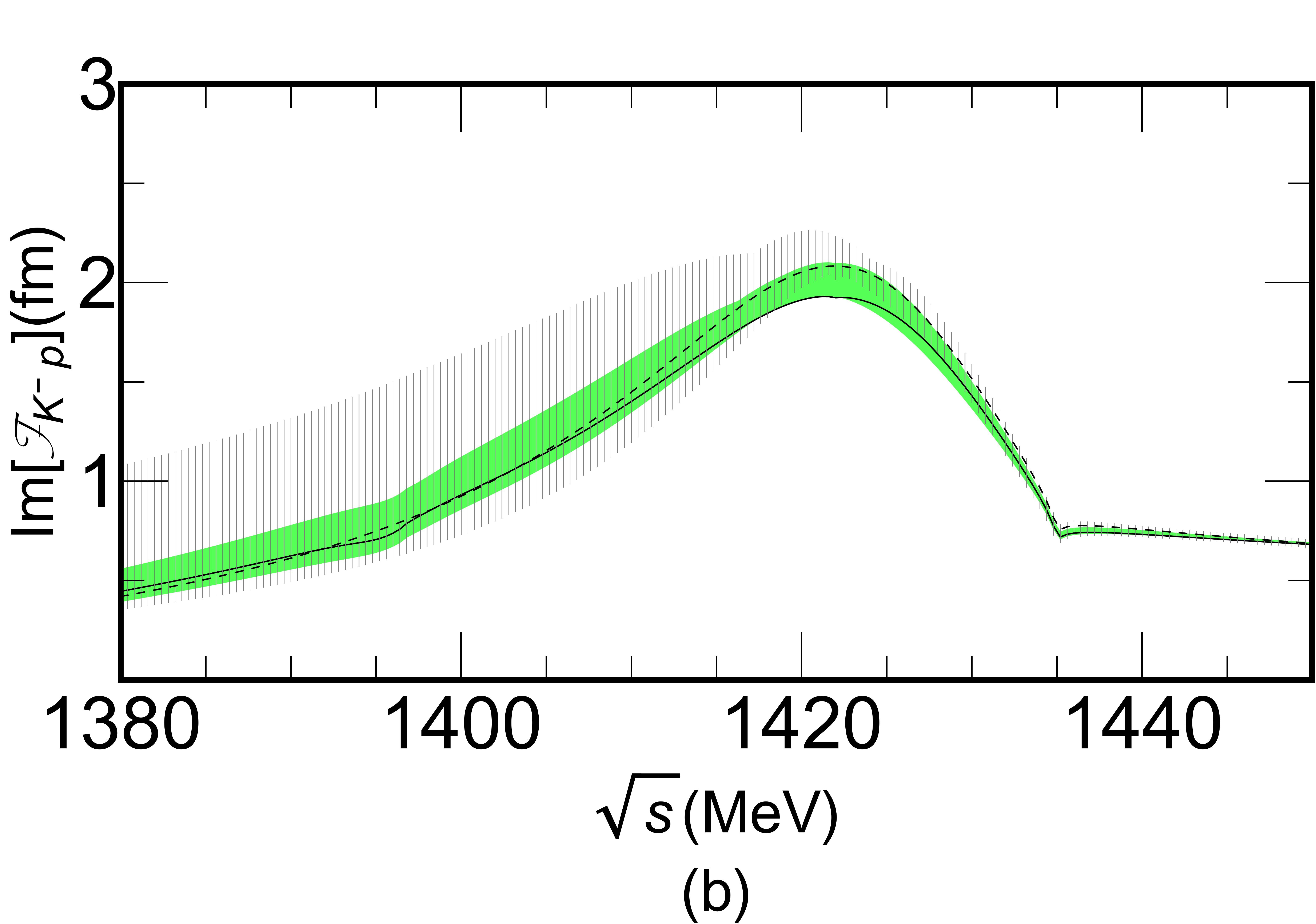}
\caption{Real (right) and imaginary (left) parts of the $\mathcal{F}$ function for the process $K^- p\to K^-p$ for Fit I (shadowed region) and Fit II (region filled with vertical lines). The solid (dashed) line represents the result of $\mathcal{F}$ associated with the minimum $\chi^2_\text{d.o.f}$ found for Fit I (Fit II).}\label{TKmpb}
\end{figure}

In Fig.~\ref{TKmpb} we show the function $\mathcal{F}(\sqrt{s})$, which is defined as
\begin{align}
\mathcal{F}_{i}(\sqrt{s})=-\frac{M_i}{4\pi\sqrt{s}}T_{ii}(\sqrt{s}),
\end{align}
with $M_i$ being the baryon mass of channel $i$, for the $K^-p$ channel. This information is relevant since the processes $K^-p\to K^- p$ in the energy region around the $\Lambda(1405)$ plays an important role when describing the absorption of $K^-$ in nuclear surfaces~\cite{Bloom:1969as}. The presence of the lighter $\Lambda(1405)$ pole extends the region of interest of the $\bar{K}N$ scattering amplitude for the antikaon self-energy in the nuclear medium toward lower energies. As can be seen, although the two sets of solutions (Fit I) and (Fit II) produce compatible results below the threshold, the uncertainty associated with the Fit II is bigger. 

For completeness, we give in Table~\ref{scat} the $K^-p$ scattering length, determined from Eq.~(\ref{aeq}), as well as the scattering lengths associated with the $\bar K N$ system in isospins $I=0$ and $I=1$, respectively. The value found for the $K^-p$ scattering length is in agreement with the one obtained using directly the SIDDHARTA data, $a_{K^-p}=(-0.65\pm0.10)+i\,(0.81\pm 0.15)$ fm, by means of Eq.~(\ref{1s_energy}), and with the result of Ref.~\cite{Iwasaki:1997wf} from Kaonic hydrogen $x$ rays, $a_{K^-p}=(-0.78\pm0.15\pm0.03)+i\,(0.49\pm 0.25\pm0.12)$ fm.
\begin{table}[h!]
\caption{Scattering lengths for $K^-p$ and $\bar K N$ in isospin 0 and 1, respectively (all units are in \text{fm}). }\label{scat}
\begin{tabular}{c|c|c}
&\text{Fit I}&\text{Fit II}\\
\hline
$a_{K^-p}$&$-0.74^{+0.01}_{-0.02}+i\,0.69^{+0.02}_{-0.01}$&$-0.74^{+0.07}_{-0.02}+i\,0.73^{+0.03}_{-0.08}$\\
$a^0_{\bar KN}$&$-1.58^{+0.03}_{-0.03}+i\,0.87^{+0.02}_{-0.03}$&$-1.60^{+0.03}_{-0.01}+i\,0.89^{+0.04}_{-0.13}$\\
$a^1_{\bar K N}$&$0.09^{+0.02}_{-0.02}+i\,0.50^{+0.04}_{-0.02}$&$0.12^{+0.10}_{-0.04}+i\,0.55^{+0.02}_{-0.04}$
\end{tabular}
\end{table}
\section{Summary and outlook}
A simultaneous fit to several relevant data has been made to study hyperon resonances. Low-lying  hyperon resonances have been studied earlier in several works, by solving pseudoscalar-baryon coupled-channel scattering equations. We have included both pseudoscalar- and  vector-baryon dynamics and find that the properties of the widely known hyperons, like, $\Lambda(1405)$, are well reproduced. The formalism used in the previous work on this topic~\cite{Khemchandani:2012ur} has been extended by including $s$- and $u$-channel diagrams to study pseudoscalar-baryon interactions. We find that an isospin 1 state, around 1400 MeV, also exists, though it is not clear if it is related to one or two poles in the complex plane. The data fitted in the present work are related to the production of pseudoscalar-baryon channels. Still the cross sections at somewhat higher energies are found to follow the data, in one of the two fits obtained in the present work. Thus, hyperons resonances with higher masses have also been studied. The present work can further be improved by considering data on reactions with vector-baryon as final states and by including decuplet baryons in our formalism.

\section*{Acknowledgments}\vspace{-0.5cm}
The authors sincerely thank Prof. Eulogio Oset for reading the manuscript and giving useful suggestions. K.P.K. and A.M.T.  gratefully acknowledge the financial support received from FAPESP (under  Grant No. 2012/50984-4) and CNPq (under Grant No. 310759/2016-1 and No. 311524/2016-8). J.A.O. thanks partial financial support from the MINECO (Spain) and FEDER (EU) Grant No. FPA2016-77313-P. 

\clearpage
\appendix*
\section{Isospin coefficients of different pseudoscalar-baryon amplitudes}
The amplitudes for the contact interaction and the $s$-channel diagram, given in  Eqs.~(\ref{Eq:WT}) and (\ref{Eq:s}), can be projected on $s$-wave to obtain,
\begin{align}
V_{\rm cont}^{L=0}& (i\to j) = -\frac{1}{4f_P^2} \sqrt{\frac{M_i + E_i}{2M_i}}\sqrt{\frac{M_j + E_j}{2M_j}}  \mathcal A_{ij} \Biggl[ \left(2\sqrt{s}-M_i -M_j \right) \Biggr],\label{Eq:WT_swave}\\\nonumber\\
V_s^{L=0}& (i\to j)  = \frac{1}{2f_P^2} \sqrt{\frac{M_i + E_i}{2M_i}}\sqrt{\frac{M_j + E_j}{2M_j}} \left(\sqrt{s}-M_i \right)\left(\sqrt{s}-M_j \right) \sum_k \frac{\mathcal B_{ij}^k}{\sqrt{s} + M_{k}}, \label{Eq:s_swave}
\end{align}
The amplitudes for the $u$-channel diagram, Eq.~(\ref{Eq:u}), is  projected on $s$-wave as follows\footnote{In the case of $|\vec p_i| = 0$ or $|\vec p_j| = 0$, $V_u^{L=0}$ can be obtained from Eq.~(\ref{Eq:u}) directly.}:
\begin{align}
V_u^{L=0}& (i\to j) = -\frac{1}{2f_P^2} \sqrt{\frac{M_i + E_i}{2M_i}}\sqrt{\frac{M_j + E_j}{2M_j}} \sum_k \mathcal C_{ij}^k \Biggl[ \sqrt{s} + M_{k}  \Biggr. \nonumber\\
& -  \frac{\left(M_i + M_k\right)\left(M_j+M_k\right) \left(\sqrt{s} + M_i + M_j - M_k \right)}{2(M_i + E_i)(M_j + E_j)} +\left( \frac{\left(M_i + M_k\right)\left(M_j+M_k\right) }{4|\vec p_i|~| \vec p_j|}\right)\nonumber\\
&\times\left( \left(\sqrt{s} - M_i - M_j + M_k \right) - \frac{s + M_k^2 - m_i^2- m_j^2 - 2 E_i E_j}{2(M_i + E_i)(M_j + E_j)} \left(\sqrt{s} + M_i + M_j - M_k \right)\right)\nonumber\\
&\times \mathcal{F}(\sqrt{s}, m_i, M_i,m_j, M_j, M_k)\Biggr],\label{Eq:u_swave}
\end{align}
where,
\begin{align}
&\mathcal{F}(\sqrt{s}, m_i, M_i,m_j, M_j, M_k)\nonumber\\
&\quad=\nonumber\left\{\begin{array}{c}-2\dfrac{s-m_i^2-m_j^2-M_k^2}{|s-m_i^2-m_j^2-M_k^2|}\text{arctg}\left[\Bigg|\dfrac{2 |\vec p_i||\vec p_j|}{s-m_i^2-m_j^2-M_k^2}\Bigg|\right],~\sqrt{s}_\text{min} < \sqrt{s} < \sqrt{s}_\text{max}\\
\\\ln\left(\dfrac{s + M_k^2 - m_i^2- m_j^2 - 2 E_i E_j - 2|\vec p_i|~| \vec p_j| }{s + M_k^2 - m_i^2- m_j^2 - 2 E_i E_j + 2|\vec p_i|~| \vec p_j|}\right),~~~~~~~~\text{otherwise}\end{array}\right.
\end{align}
where $\sqrt{s}_\text{min}=\text{min}(m_i+M_i, m_j+M_j)$, $m_i$ ($m_j$) represents the meson mass in the initial (final) state, $M_i$, $E_i$ ($M_j$, $E_j$) represent the mass and energy of the baryon in the initial (final) state, $M_k$ is the mass of the baryon exchanged in the $s$-, $u$-channels. The values for $\mathcal A_{ij}$, $\mathcal B_{ij}$, $\mathcal C_{ij}$ are given in Tables~\ref{iso0s1hdirect}, \ref{iso1s1hdirect}, \ref{iso0s1hcross}, and \ref{iso1s1hcross} for different isospins and different processes. In Eq.~(\ref{Eq:u_swave}), $|\vec p_j|$ is the center of mass momentum of the $j$th channel,
\begin{equation}
|\vec p_j| = \frac{1}{2\sqrt{s}}\left[\lambda(s,m_j^2,M_j^2)\right]^{1/2},
\end{equation}
which  becomes complex valued below the threshold.

\setcounter{table}{0}
\renewcommand{\thetable}{A\arabic{table}}
{\squeezetable \begin{table}[h!]
\caption{Coefficients for the $s$-channel amplitudes in the isospin 0 base. We indicate in the first column the exchanged particles. For example, the only nonzero contribution to an $s$-channel diagram for $\bar K N\to \bar KN$,  in the isospin 0, comes from a $\Lambda$  exchange.\\}\label{iso0s1hdirect}
\begin{ruledtabular}
\begin{tabular}{c|c|c|c|c|c}
&\multicolumn{2}{c|}{$\bar K N$}&$K\Xi$&$\pi\Sigma$&$\eta\Lambda$\\
\hline
\multirow{4}{*}{$\bar KN$}&$\Sigma$&0&0&0&0\\
&$\Lambda$&$\frac{(D^\prime+3F^\prime)^2}{3}$&$3F^{\prime\,2}-\frac{D^{\prime\,2}}{3}$&$\sqrt{\frac{2}{3}}D^\prime(D^\prime+3F^\prime)$&$\frac{\sqrt{2}}{3}D^\prime(D^\prime+3F^\prime)$\\
&$N$&0&0&0&0\\
&$\Xi$&0&0&0&0\\
\hline
\multirow{4}{*}{$\bar K\Xi$}&$\Sigma$&0&0&0&0\\
&$\Lambda$&$3F^{\prime\,2}-\frac{D^{\prime\,2}}{3}$&$\frac{(D^\prime-3F^\prime)^2}{3}$&$-\sqrt{\frac{2}{3}}D^\prime(D^\prime-3F^\prime)$&$-\frac{\sqrt{2}}{3}D^\prime(D^\prime-3F^\prime)$\\
&$N$&0&0&0&0\\
&$\Xi$&0&0&0&0\\
\hline
\multirow{4}{*}{$\pi\Sigma$}&$\Sigma$&0&0&0&0\\
&$\Lambda$&$\sqrt{\frac{2}{3}}D^\prime(D^\prime+3F^\prime)$&$-\sqrt{\frac{2}{3}}D^\prime(D^\prime-3F^\prime)$&$2D^{\prime\,2}$&$\frac{2D^{\prime\,2}}{\sqrt{3}}$\\
&$N$&0&0&0&0\\
&$\Xi$&0&0&0&0\\
\hline
\multirow{4}{*}{$\eta\Lambda$}&$\Sigma$&0&0&0&0\\
&$\Lambda$&$\frac{\sqrt{2}}{3}D^\prime(D^\prime+3F^\prime)$&$-\frac{\sqrt{2}}{3}D^\prime(D^\prime-3F^\prime)$&$\frac{2D^{\prime\,2}}{\sqrt{3}}$&$\frac{2D^{\prime\,2}}{3}$\\
&$N$&0&0&0&0\\
&$\Xi$&0&0&0&0\\
\end{tabular}
\end{ruledtabular}
\end{table}}

{\squeezetable\begin{table}
\caption{Coefficients for the $s$-channel amplitudes in the isospin 1 base. We indicate in the first column the exchanged particles. For example, the only nonzero contribution to a $s$-channel diagram for $\bar K N\to \bar KN$,  in isospin 1, comes from a $\Sigma$  exchange.\\}\label{iso1s1hdirect}
\begin{ruledtabular}
\begin{tabular}{c|c|c|c|c|c|c}
&\multicolumn{2}{c|}{$\bar K N$}&$K\Xi$&$\pi\Sigma$&$\pi\Lambda$&$\eta\Sigma$\\
\hline
\multirow{4}{*}{$\bar KN$}&$\Sigma$&$(D^\prime-F^\prime)^2$&$D^{\prime\,2}-F^{\prime\,2}$&$2F^\prime(D^\prime-F^\prime)$&$-\sqrt{\frac{2}{3}}D^\prime(D^\prime-F^\prime)$&$-\sqrt{\frac{2}{3}}D^\prime(D^\prime-F^\prime)$\\
&$\Lambda$&0&0&0&0&0\\
&$N$&0&0&0&0&0\\
&$\Xi$&0&0&0&0&0\\
\hline
\multirow{4}{*}{$\bar K\Xi$}&$\Sigma$&$D^{\prime\,2}-F^{\prime\,2}$&$(D^\prime+F^\prime)^2$&$2F^\prime(D^\prime+F^\prime)$&$-\sqrt{\frac{2}{3}}D^\prime(D^\prime+F^\prime)$&$-\sqrt{\frac{2}{3}}D^\prime(D^\prime+F^\prime)$\\
&$\Lambda$&0&0&0&0&0\\
&$N$&0&0&0&0&0\\
&$\Xi$&0&0&0&0&0\\
\hline
\multirow{4}{*}{$\pi\Sigma$}&$\Sigma$&$2F^\prime(D^\prime-F^\prime)$&$2F^\prime(D^\prime+F^\prime)$&$4F^{\prime\,2}$&$-2\sqrt{\frac{2}{3}}D^\prime F^\prime$&$-2\sqrt{\frac{2}{3}}D^\prime F^\prime$\\
&$\Lambda$&0&0&0&0&0\\
&$N$&0&0&0&0&0\\
&$\Xi$&0&0&0&0&0\\
\hline
\multirow{4}{*}{$\pi\Lambda$}&$\Sigma$&$-\sqrt{\frac{2}{3}}D^\prime(D^\prime-F^\prime)$&$-\sqrt{\frac{2}{3}}D^\prime(D^\prime+F^\prime)$&$-2\sqrt{\frac{2}{3}}D^\prime F^\prime$&$\frac{2D^{\prime\,2}}{3}$&$\frac{2D^{\prime\,2}}{3}$\\
&$\Lambda$&0&0&0&0&0\\
&$N$&0&0&0&0&0\\
&$\Xi$&0&0&0&0&0\\
\hline
\multirow{4}{*}{$\eta\Sigma$}&$\Sigma$&$-\sqrt{\frac{2}{3}}D^\prime(D^\prime-F^\prime)$&$-\sqrt{\frac{2}{3}}D^\prime(D^\prime+F^\prime)$&$-2\sqrt{\frac{2}{3}}D^\prime F^\prime$&$\frac{2D^{\prime\,2}}{3}$&$\frac{2D^{\prime\,2}}{3}$\\
&$\Lambda$&0&0&0&0&0\\
&$N$&0&0&0&0&0\\
&$\Xi$&0&0&0&0&0\\
\end{tabular}
\end{ruledtabular}
\end{table}}

{\squeezetable \begin{table}
\caption{Coefficients for the $u$-channel  amplitudes in the isospin 0 base. We indicate in the first column the exchanged particles. For example, a $\Sigma$ and  a $\Lambda$  exchange in the $u$-channel give  nonzero contributions to the process $\bar K N\to \bar K\Xi$,  in isospin 0.\\}\label{iso0s1hcross}
\begin{ruledtabular}
\begin{tabular}{c|c|c|c|c|c}
&\multicolumn{2}{c|}{$\bar K N$}&$K\Xi$&$\pi\Sigma$&$\eta\Lambda$\\
\hline
\multirow{4}{*}{$\bar KN$}&$\Sigma$&0&$-\frac{3}{2}(D^{\prime\,2}-F^{\prime\,2})$&0&0\\
&$\Lambda$&0&$-\frac{1}{6}(D^{\prime\,2}-9F^{\prime\,2})$&0&0\\
&$N$&0&0&$-\sqrt{\frac{3}{2}}(D^{\prime\,2}-F^{\prime\,2})$&$\frac{D^{\prime\,2}-9F^{\prime\,2}}{3\sqrt{2}}$\\
&$\Xi$&0&0&0&0\\
\hline
\multirow{4}{*}{$\bar K\Xi$}&$\Sigma$&$-\frac{3}{2}(D^{\prime\,2}-F^{\prime\,2})$&0&0&0\\
&$\Lambda$&$-\frac{1}{6}(D^{\prime\,2}-9F^{\prime\,2})$&0&0&0\\
&$N$&0&0&0&0\\
&$\Xi$&0&0&$\sqrt{\frac{3}{2}}(D^{\prime\,2}-F^{\prime\,2})$&$-\frac{D^{\prime\,2}-9F^{\prime\,2}}{3\sqrt{2}}$\\
\hline
\multirow{4}{*}{$\pi\Sigma$}&$\Sigma$&0&0&$-4F^{\prime\,2}$&$-\frac{2D^{\prime\,2}}{\sqrt{3}}$\\
&$\Lambda$&0&0&$\frac{2D^{\prime\,2}}{3}$&0\\
&$N$&$-\sqrt{\frac{3}{2}}(D^{\prime\,2}-F^{\prime\,2})$&0&0&0\\
&$\Xi$&0&$\sqrt{\frac{3}{2}}(D^{\prime\,2}-F^{\prime\,2})$&0&0\\
\hline
\multirow{4}{*}{$\eta\Lambda$}&$\Sigma$&0&0&$-\frac{2D^{\prime\,2}}{\sqrt{3}}$&0\\
&$\Lambda$&0&0&0&$\frac{2D^{\prime\,2}}{3}$\\
&$N$&$\frac{D^{\prime\,2}-9F^{\prime\,2}}{3\sqrt{2}}$&0&0&0\\
&$\Xi$&0&$-\frac{D^{\prime\,2}-9F^{\prime\,2}}{3\sqrt{2}}$&0&0\\
\end{tabular}
\end{ruledtabular}
\end{table}}

{\squeezetable \begin{table}
\caption{Coefficients for the $u$-channel amplitudes in the isospin 1 base. We indicate in the first column the exchanged particles. For example, a $\Sigma$ and a $\Lambda$ in the $u$-channel give nonzero contributions to the process $\bar K N\to \bar K\Xi$,  in isospin 1.\\}\label{iso1s1hcross}
\begin{ruledtabular}
\begin{tabular}{c|c|c|c|c|c|c}
&\multicolumn{2}{c|}{$\bar K N$}&$K\Xi$&$\pi\Sigma$&$\pi\Lambda$&$\eta\Sigma$\\
\hline
\multirow{4}{*}{$\bar KN$}&$\Sigma$&0&$-\frac{D^{\prime\,2}-F^{\prime\,2}}{2}$&0&0&0\\
&$\Lambda$&0&$\frac{(D^{\prime\,2}-9F^{\prime\,2})}{6}$&0&0&0\\
&$N$&0&0&$D^{\prime\,2}-F^{\prime\,2}$&$\frac{(D^\prime+F^\prime)(D^\prime+3F^\prime)}{\sqrt{6}}$&$\frac{(D^\prime-F^\prime)(D^\prime-3F^\prime)}{\sqrt{6}}$\\
&$\Xi$&0&0&0&0&0\\
\hline
\multirow{4}{*}{$\bar K\Xi$}&$\Sigma$&$-\frac{D^{\prime\,2}-F^{\prime\,2}}{2}$&0&0&0&0\\
&$\Lambda$&$\frac{(D^{\prime\,2}-9F^{\prime\,2})}{6}$&0&0&0&0\\
&$N$&0&0&0&0&0\\
&$\Xi$&0&0&$F^{\prime\,2}-D^{\prime\,2}$&$\frac{(D^\prime-F^\prime)(D^\prime-3F^\prime)}{\sqrt{6}}$&$\frac{(D^\prime+F^\prime)(D^\prime+3F^\prime)}{\sqrt{6}}$\\
\hline
\multirow{4}{*}{$\pi\Sigma$}&$\Sigma$&0&0&$2F^{\prime\,2}$&$2\sqrt{\frac{2}{3}}D^\prime F^\prime$&$-2\sqrt{\frac{2}{3}}D^\prime F^\prime$\\
&$\Lambda$&0&0&$-\frac{2D^{\prime\,2}}{3}$&0&0\\
&$N$&$D^{\prime\,2}-F^{\prime\,2}$&0&0&0&0\\
&$\Xi$&0&$F^{\prime\,2}-D^{\prime\,2}$&0&0&0\\
\hline
\multirow{4}{*}{$\pi\Lambda$}&$\Sigma$&0&0&$2\sqrt{\frac{2}{3}}D^\prime F^\prime$&$\frac{2D^{\prime\,2}}{3}$&0\\
&$\Lambda$&0&0&0&0&$-\frac{2D^{\prime\,2}}{3}$\\
&$N$&$\frac{(D^\prime+F^\prime)(D^\prime+3F^\prime)}{\sqrt{6}}$&0&0&0&0\\
&$\Xi$&0&$\frac{(D^\prime-F^\prime)(D^\prime-3F^\prime)}{\sqrt{6}}$&0&0&0\\
\hline
\multirow{4}{*}{$\eta\Sigma$}&$\Sigma$&0&0&$-2\sqrt{\frac{2}{3}}D^\prime F^\prime$&0&$\frac{2D^{\prime\,2}}{3}$\\
&$\Lambda$&0&0&0&$-\frac{2D^{\prime\,2}}{3}$&0\\
&$N$&$\frac{(D^\prime-F^\prime)(D^\prime-3F^\prime)}{\sqrt{6}}$&0&0&0&0\\
&$\Xi$&0&$\frac{(D^\prime+F^\prime)(D^\prime+3F^\prime)}{\sqrt{6}}$&0&0&0\\
\end{tabular}
\end{ruledtabular}
\end{table}}

\end{document}